\documentclass[journal]{IEEEtran}
\usepackage{float}
\usepackage{bbm}
\usepackage{footmisc}

\usepackage{multirow}
\usepackage{amssymb}
\usepackage{amsmath}
\usepackage{mathrsfs}
\usepackage{amsfonts}
\usepackage{graphicx}
\usepackage{tabularx}
\usepackage{epstopdf}
\usepackage{microtype}
\usepackage{stfloats}
\usepackage[numbers,sort&compress]{natbib}
\usepackage[noend]{algpseudocode}
\usepackage{algorithmicx,algorithm}
\usepackage{color}

\newtheorem{theorem}{Theorem}
\newtheorem{corollary}{Corollary}
\newtheorem{lemma}{Lemma}
\newtheorem{remark}{Remark}

\newtheorem{proposition}{Proposition}
\renewcommand{\IEEEQED}{\IEEEQEDopen}
\makeatletter

\newcommand{\Rmnum}[1]{\expandafter\@slowromancap\romannumeral #1@}
\IEEEoverridecommandlockouts
\title{Energy-Aware Offloading in Time-Sensitive Networks with Mobile Edge Computing}
\author{Mingxiong~Zhao,~\IEEEmembership{Member,~IEEE,}
	Jun-Jie~Yu,
	Wen-Tao~Li,
	Di~Liu,	
	Shaowen~Yao,\\
	Wei~Feng,
    Changyang~She,~\IEEEmembership{Member,~IEEE,}
	Tony Q. S. Quek,~\IEEEmembership{Fellow,~IEEE}\\
	\thanks{M. Zhao, J. Yu, W. Li, D. Liu and S. Yao are with National Pilot School of Software, Yunnan University, Kunming, China. (Email: mx\_zhao@ynu.edu.cn, yjj19951118@gmail.com, lwt@mail.ynu.edu.cn, dliu@ynu.edu.cn, yaosw@ynu.edu.cn.)}
	\thanks{W. Feng is with School of Communication Engineering, Hangzhou Dianzi University, Hangzhou, China. Email: fengwei@hdu.edu.cn.}
	\thanks{C. She is with the School of Electrical and Information Engineering, University of Sydney, Sydney, NSW 2006, Australia. Email: shechangyang@gmail.com.}
	\thanks{T. Q. S. Quek is with Singapore University of Technology and Design,
		Singapore 487372. E-mail: tonyquek@sutd.edu.sg.}}
\begin{document}
	\maketitle
\begin{abstract}
	Mobile Edge Computing (MEC) enables rich services in close proximity to the end users to provide high quality of experience (QoE) and contributes to energy conservation compared with local computing, but results in increased communication latency. In this paper, we investigate how to jointly optimize task offloading and resource allocation to minimize the energy consumption in an orthogonal frequency division multiple access-based MEC networks, where the time-sensitive tasks can be processed at both local users and MEC server via partial offloading. Since the optimization variables of the problem are strongly coupled, we first decompose the orignal problem into three subproblems named as offloading selection ($\mathbf{P_O}$), transmission power optimization ($\mathbf{P_T}$), and subcarriers and computing resource allocation ($\mathbf{P_S}$), and then propose an iterative algorithm to deal with them in a sequence. To be specific, we derive the closed-form solution for $\mathbf{P_O}$, employ the equivalent parametric convex programming to cope with the objective function which is in the form of sum of ratios in $\mathbf{P_T}$, and deal with $\mathbf{P_S}$ by an alternating way in the dual domain due to its NP-hardness. Simulation results demonstrate that the proposed algorithm outperforms the existing schemes.
\end{abstract}
\section{Introduction}
With the rapid development of Internet of Things (IoT), the demand for high-speed, ultra-low latency and dynamically configurable computing resources at the edge of networks is extensively growing \cite{mu2018joint}. The increasingly high computation loads and critical latency requirements by the applications (e.g., virtual reality/augmented reality (AR/VR), online games and drones) brings significant challenges to the IoT devices with limited computational capability and battery lifetime. To deal with such a thorny problem, the new technology named as Mobile Edge Computing (MEC) has been envisioned as a potential approach to push the cloud services down to the vicinity of IoT devices, and process the offloaded computation-intensive or energy-consuming tasks from IoT devices in recent years \cite{Porambage2018,yang2018multi}.

As an effective method to liberate IoT devices from computation-intensive computing workloads, MEC can efficiently reduce the energy consumption of IoT devices, and thus has been considered as a promising architecture for the scenarios with energy-constrained IoT devices \cite{chen2019ENe,cheng2018energy,dai2018joint,liu2019dynamic}. To be specific, the authors in \cite{chen2019ENe} described the task offloading of MEC system as a stochastic optimization problem, which was further translated into
a deterministic optimization problem, and proposed a dynamic offloading algorithm to minimize system energy consumption. In \cite{cheng2018energy} and \cite{dai2018joint}, wireless resource allocation and user association were optimized, respectively, to minimize the total energy consumptions of MEC systems. Moreover, the authors in \cite{liu2019dynamic} aimed to minimize users' power consumption while trading off the allocated resources for local computation and task offloading. 
\raggedbottom

However, offloading energy-consuming workloads to MEC servers also invokes extra latency which significantly affects the quality of experience (QoE) of users, and thus cannot be ignored in the system design. As one of the important metrics to measure the performance of MEC network, time sensitivity\footnote{Specifically, applications in industrial IoT (I-IoT) or military industry generally have rigorous criteria on task processing time and require the tasks precisely calculated and completed within certain time interval \cite{Parvez2018}.} has triggered more and more research interests and has been investigated in the literature \cite{ren2018latency,ren2019collaborative,Changyang2019IoT,cui2018joint,kuang2019partial}. To be specific, the authors investigated the latency-minimization problem in a multi-user time-division multiple access MEC offloading system \cite{ren2018latency}, and then minimized the weighted-sum latency of all mobile devices via the collaboration between cloud computing and edge computing \cite{ren2019collaborative}. Furthermore, the authors in \cite{Changyang2019IoT} derived the distribution of processing delay of mission-critical IoTs in MEC networks and further optimized user association, task offloading and bandwidth allocation. Meanwhile, some pioneering works considered the energy consumption minimization with the QoE requirement of time-sensitive computation tasks \cite{cui2018joint,kuang2019partial}. To satisfy user demands of various IoT applications, the authors in \cite{cui2018joint} found a tradeoff between the energy consumption and latency, and formalized the problem into a constrained multi-objective optimization problem. At the same time, the authors in \cite{kuang2019partial} minimized the weighted sum of the execution delay and energy consumption while guaranteeing the transmission power constraint of IoT devices based on partial offloading.
\raggedbottom

To further improve the utilization of radio resources, Orthogonal Frequency-Division Multiple Access (OFDMA) has been widely applied to the MEC system for various objectives in MEC system, such as profit or utility\footnote{In general, the profit or utility implies the revenue from computation offloading for IoT devices, or the reward for MEC severs through helping computation.} maximization \cite{kim2018optimal,paymard2019task}, computation efficiency maximization \cite{wu2019computation}, delay minimization \cite{li2017joint}, the energy consumption minimization with or without the requirement of computation latency
%cheng2018energy
\cite{zhang2017joint,wen2019joint,khalili2019joint,yang2019energy,you2016energy}, and the energy-latency tradeoff \cite{zhang2017energy}. To be specific, the utilities of the mobile users and the edge clouds were optimized while considering the price of computation capacity in \cite{kim2018optimal} using game theory. The authors in \cite{paymard2019task} propsosed a priority-based task scheduling policy and jointly optimized the computation and communication resource to maximize the profit of mobile network operator while satisfying users' quality of service (QoS). Furthermore, the authors in \cite{wu2019computation} investigated the weighted sum computation efficiency maximization problems for both partial and binary offloading mode. To minimize the maximum delay of each mobile device, the authors in \cite{li2017joint} considered a partial offloading scheme and developed a heuristic algorithm to jointly optimize the subcarrier and power allocation. At the meantime, the authors in \cite{zhang2017joint} minimized the energy consumption and monetary cost, i.e., communication and computation cost from mobile terminals' perspective. An energy-consumption minimization problem was considered in \cite{wen2019joint} according to formulate a joint uplink/downlink sub-channel, bit and time allocation problem, meanwhile, QoS was introduced to minimize the energy consumption of a multi-cell MEC network in \cite{khalili2019joint}. Moreover, the authors proposed an energy-efficient joint offloading and wireless resource allocation strategy for delay-critical applications to minimize the total (or weighted-sum) energy consumption of systems
%cheng2018energy,
\cite{you2016energy} or of the mobile devices \cite{yang2019energy}, and the tradeoff between energy consumption and sensitive latency was further considered to design the energy-aware offloading scheme in \cite{zhang2017energy}.

%However, the pioneering works \cite{wen2019joint,you2016energy, zhang2017energy,cheng2018Ene,yang2019energy} investigated binary offloading without the consideration of partial offloading. More specifically, the authors in \cite{wen2019joint} assumed that each task must be uploaded to serving node for processing, which can be treated as a special case of binary offloading. Meanwhile, the authors in \cite{cheng2018Ene,yang2019energy} simplified the binary offloading strategy according to a threshold value given by local computing time, and thus they did not take the QoE requirement of time-sensitive computation tasks into more detailed consideration, which was further investigated in \cite{you2016energy, zhang2017energy}. Motivated by the aforementioned issues, we consider partial offloading where mobile date can be computed at both local devices and MEC server, and investigate an OFDMA-based MEC network in this paper. Our designed framework aims to minimize the total energy consumption of the whole network, where offloading ratio, transmission power, computation capability and subcarriers are jointly optimized to satisfy the QoE requirement of time-sensitive computation tasks of users. The main contributions of this paper are summarized as follows:
%
However, the inspiring works \cite{wen2019joint,you2016energy, zhang2017energy,yang2019energy} only investigated binary offloading without the consideration of partial offloading, which can flexibly allocate resources for computation offloading and local computing, and thus achieve better performace \cite{wu2019computation,zhou2018computation}, although the QoE requirement of time-sensitive computation tasks was taken into account \cite{wen2019joint,you2016energy, zhang2017energy}. Motivated by the aforementioned issues, we consider partial offloading where mobile date can be computed at both local users and MEC server, and investigate an OFDMA-based MEC network in this paper. Our designed framework aims to minimize the total energy consumption of the whole network, where offloading ratio, transmission power, computation capability and subcarriers are jointly optimized to satisfy the QoE requirement of time-sensitive computation tasks of users. The main contributions of this paper are summarized as follows:

1) We jointly optimize offloading ratio, transmission power, computation capability assignment, and subcarrier allocation to minimize the total energy consumption of the OFDMA-based MEC network with the QoE requirement of time-sensitive computation tasks.

2) We decompose the original non-convex and strongly coupled problem into three subproblems with respect to offloading ratio selection $\mathbf{P_O}$, transmission power optimization $\mathbf{P_T}$, subcarriers and computing resource allocation $\mathbf{P_S}$, respectively, and optimize them in a sequence according to Block Coordinate Descent (BCD) method.

3) To solve the subproblems, we first derive the closed-form solution of the offloading ratios for $\mathbf{P_O}$. Secondly, we apply the equivalent parametric convex programming to tackle the form of sum of ratios in $\mathbf{P_T}$, and then assign the transmission power among subcarriers with the help of auxiliary variables. Lastly, since $\mathbf{P_S}$ is a mixed integer nonlinear programming (MINLP) problem, we assign computation capability and allocate subcarriers to each user, and further optimize them in an alternating manner in dual domain.

{\spaceskip=0.2em\relax The rest of paper is organized as follows.} The system model and problem formulation are introduced in Section \ref{2}. In Section \ref{3}, offloading and resource allocation strategy is proposed.  We further assume equal power allocation to optimize the transmission power in Section \ref{OP2-EPA}. Numerical results {\spaceskip=0.3em\relax are presented in Section \ref{5}, and conclusions are drawn in Section \ref{6}.}

\section{System model and problem formulation}\label{2}
\begin{figure}  
	\centering  
	\includegraphics[height=6cm]{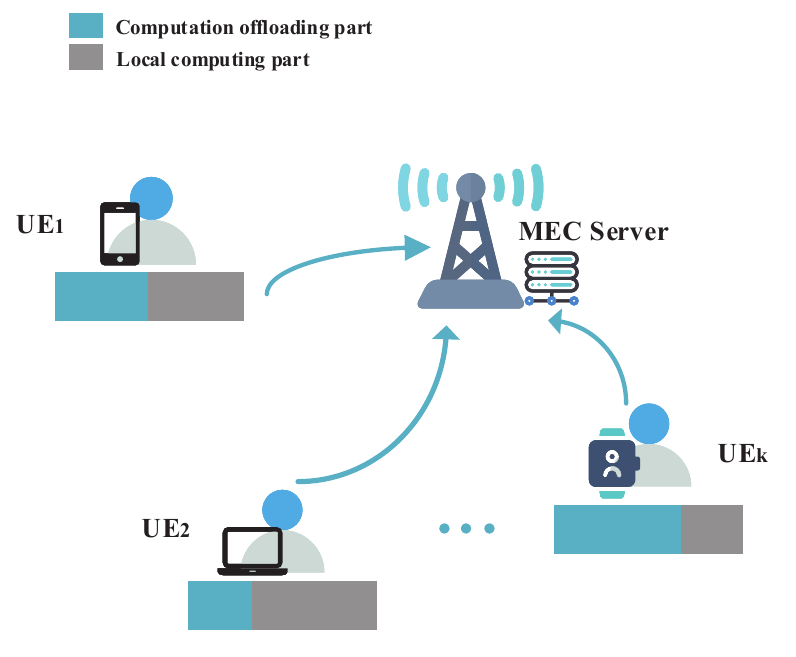}\\
	\caption{Illustration of our system model.}  
	\label{fig_model} 
\end{figure}

We consider an OFDMA-based MEC system with $K$ users and one base-station (BS) integrated with an MEC server to execute the time-sensitive computation tasks. All nodes are equipped with a single antenna as shown in Fig.\ref{fig_model}. Denote $\mathcal{K} \triangleq \{1,2, \cdots, K\}$ as the set of users, and let  $\mathcal{N}\triangleq \{1,2, \cdots, N\}$ be the index for multiple orthogonal subcarriers, each of which has bandwidth $B$ and can be assigned to only one user. In this system, we assume that user $k$ has a task described by a tuple of four parameters $\{R_k, c_k, \lambda_k, t_k\}$, where $R_{k}$ indicates the amount of input data to be processed, $c_{k}$ reprsesents the number of CPU cycles for computing 1-bit of input data, $\lambda_k\in[0,1]$ is the proportion of $R_k$ offloading to MEC, while the rest $(1-\lambda_k)R_k$ bits are processed by its local CPU, and $t_k$ is the maximum tolerable latency. In this paper, it is assumed that the maximum tolerable latency for user $k$, $t_k$ is shorter than the channel coherence time, such that the wireless channels remain constant during a time slot with length $T$, i.e., $t_k\leq T, \forall k$, but can vary from time to time. The local CPU frequency of user $k$ is characterized by $f_k$, and $f_{k,m}$ is the computing resource allocated to user $k$ from MEC server, where both of them are measured by the number of CPU cycles per second. Herein, a practical constraint that the total computing resources allocated to all the associated users must not excess the server’s computing capacity $F$, is given by $\sum_{k\in\mathcal{K}}f_{k,m}\leq F$. 

The notations used in the paper are summarized in TABLE \uppercase\expandafter{\romannumeral1}. In the following, the latency and the energy consumption of user $k$ for our considered system are given in details.
\subsection{Latency}
\subsubsection{Local Computing at Users} 
Consider the local computing for executing the residual $(1-\lambda_k)R_k$ input bits  at user $k$, the time consumption for local computing at user $k$ is
\begin{equation}\label{local time}
t_{k,l} =\frac{c_{k}\left(1-\lambda_{k}\right)R_{k}}{f_{k}}.
\end{equation}
\subsubsection{Computation Offloading} 
According to the OFDMA mechanism, the inter-interference is ignored in virtue of the exclusive subcarrier allocation. Therefore, the aggregated transmission rate to offload $\lambda_kR_k$ input bits from user $k$ to MEC server is expressed as
\begin{equation}\label{r_k}
r_k =B\sum_{n\in\mathcal{N}}x_{k,n}\log_2\left(1+p_{k,n}\tilde{g}_{k,n}\right),
\end{equation}
where $\tilde{g}_{k,n}\triangleq\frac{g_{k,n}}{\sigma_n^2}$, $g_{k,n}$ and $\sigma_n^2$ are the channel gain between user $k$ and BS, and the variance of the additive white Gaussian noise at BS on subcarrier $n$, respectively, where we set $\sigma_n^2=\sigma^2, \forall n$. Denote $p_{k,n}$ as the transmission power of the $k$-th user on the $n$-th subcarrier, which can be allocated by the user and constrained by the maximum transmission power $p_k^\text{max}$. Meanwhile, denote $x_{k,n}$ as the channel allocation indicator, specifically $x_{k,n}=1$ means that subcarrier $n$ is assigned to user $k$, otherwise $x_{k,n}=0$. In the following, $\tilde{g}_{k,n}$ can be treated as the channel gain for brevity.
% Table generated by Excel2LaTeX from sheet 'Sheet1'
\begin{table*}[htbp]
	\centering
	\caption{\textbf{SUMMARY OF NOTATIONS}}
	\begin{tabular}{||m{4.055em}|m{13em}||m{4.055em}|m{12.945em}||m{4.055em}|m{12.39em}|}
		%居中
		\hline
		\multicolumn{1}{|l|}{\textbf{Notaion}} & \multicolumn{1}{l||}{\textbf{Definition}} & \multicolumn{1}{l|}{\textbf{Notation}} & \multicolumn{1}{l||}{\textbf{Definition}} & \multicolumn{1}{l|}{\textbf{Notation}} & \multicolumn{1}{l|}{\textbf{Definition}} \\
		\hline
		\hline
		\multicolumn{1}{|p{4.055em}|}{$a_k$} &  Auxiliary variable & $g_{k,n}$ & The channel gain between user $k$ and BS & $T$ & The length of a time slot \\
		\hline
		\multicolumn{1}{|p{4.055em}|}{$a_k^i$} & $a_k$ in the $i$-th iteration &   $\tilde{g}_{k,n}$ & Related variants of $g_{k,n}$ & $t_{k,u}$ & Delay for transmitting the data of user $k$\\
		\hline
		\multicolumn{1}{|p{4.055em}|}{$B$} & Subcarrier bandwidth & $h_{k}$ & Equation \eqref{OP-PH} substitution function & $t_{k,l}$ & Delay of local calculation for the task of user $k$\\
		\hline
		\multicolumn{1}{|p{4.055em}|}{$b_k$} & Auxiliary variable &   $\kappa_k$ & The computation energy efficiency coefficient of user $k$ & $t_{k,m}$ & Delay of MEC calculation for the task of user $k$\\
		\hline
		\multicolumn{1}{|p{4.055em}|}{$b_k^i$} & $b_k$ in the $i$-th iteration &    $\kappa_m$ & The computation energy efficiency coefficient of MEC & $t_{k,\text{off}}$ & Delay for completing the offloading task of user $k$\\
		\hline
		\multicolumn{1}{|p{4.055em}|}{$c_k$} & The number of CPU cycles for computing 1-bit of input data & $|N_k|$     & The size of subcarriers assigned to user $k$ & $t$     & Iteration variable in Algorithm 3 \\
		\hline
		\multicolumn{1}{|p{4.055em}|}{$d_k$} & Equation \eqref{OP-PD} substitution function & $\mathcal{N}_k$ & The set of subcarriers allocated to user $k$  & $t_\text{max}$ & Maximum number of iterations in Algorithm 3 \\
		\hline
		\multicolumn{1}{|p{4.055em}|}{$E_{k,l}$} & Energy consumed by user $k$ locally & $N$     & Number of subcarriers & $\omega$     & Function for notational brevity \\
		\hline
		\multicolumn{1}{|p{4.055em}|}{$E_{k,m}$} & Energy consumed by user $k$ in MEC calculation & $p_{k}^{\text{max}}$ & Maximum transmission power of user $k$ & $x_{k,n}$ & The channel allocation indicator \\
		\hline
		\multicolumn{1}{|p{4.055em}|}{$E_{k,\text{off}}$} & Energy consumed by user $k$ when uploading and computing at MEC & $p_{k,n}$ & Power allocated by user $k$ on subcarrier $n$ &   $\sigma_n^2$ & The variance of the additive white Gaussian noise at BS on subcarrier $n$ \\
		\hline
		\multicolumn{1}{|p{4.055em}|}{$E_{k,u}$} & Energy consumed by user $k$ to transmit data & $p_{k,n}^{r}$ & Fixed point in Algorithm 1 &   $\lambda_{k}$ & The proportion\newline of $R_k$ offloading to MEC \\
		\hline
		\multicolumn{1}{|p{4.055em}|}{$E_k$} & The energy consumed by user $k$ to complete the task & $\boldsymbol{q}$     & The Jacobian matrix of $\boldsymbol{W}$ &   $\varphi_{k}$ & User $k$ lagrange multiplier in   \eqref{3B} \\
		\hline
		\multicolumn{1}{|m{4.055em}|}{$F$} & The CPU frequency of MEC sever & $\boldsymbol{W}$ & Auxiliary equation related to \eqref{OP2-UAb} and \eqref{OP2-UAa} &   $\vartheta_{k}$ & User $k$ lagrange multiplier in   \eqref{3B} \\
		\hline
		\multicolumn{1}{|p{4.055em}|}{$f_k$} & The CPU frequency of user $k$ & $r_{k}$ & User $k$'s transmission rate &   $\iota_{k}$ & The stepsizes corresponding to the related dual variable during iterations  in  \eqref{3B} \\
	\hline
	\multicolumn{1}{|p{4.055em}|}{$f_{k,m}$} & Calculation frequency assigned by MEC to user $k$ & $R_{k}$ & The amount of data for the task of user $k$ &   $\nu_{k}$ & The stepsizes corresponding to the related dual variable during iterations  in   \eqref{3B} \\
	\hline
	\multicolumn{1}{|p{4.055em}|}{$f_{k,m}^{\text{UB}}$} & Upper bound of $f_{k,m}$ in Algorithm 2 & $t_{k}$ & Delay for completing the task of user $k$ &  $K$     & Number of users  \\
	\hline
	\multicolumn{1}{|p{4.055em}|}{$f_{k,m}^{\text{LB}}$} & Lower bound of $f_{k,m}$ in Algorithm 2 & $l$     & Parameter for updating auxiliary variables $(a_{k}, b_{k})$&  $\mathcal{K}$ & The set of users \\
	\hline
	\multicolumn{1}{|p{4.055em}|}{$\bar{p}_k$}  & The equal power allocation for each subcarrier of user $k$ in \ref{OP2-EPA} &   $r$  & Iteration variable in Algorithm 1 &   $\tau$ & Parameter for updating auxiliary variables $(a_{k}, b_{k})$ \\
	\hline
	\multicolumn{1}{|p{4.055em}|}{$z$}  & Parameter for updating auxiliary variables $(a_{k}, b_{k})$ &   $\alpha_k$ & Lagrangian multiplier of  \eqref{OP3-L} &   $\delta_k$ & Lagrangian multiplier of  \eqref{OP3-L} \\
	\hline
	\multicolumn{1}{|p{4.055em}|}{$\phi_{k}$} & Auxiliary variable in   \eqref{3C} &   $\beta_k$ & Lagrangian multiplier of  \eqref{OP3-L} &   $\gamma$ & Lagrangian multiplier of  \eqref{OP3-L} \\
	\hline
	\multicolumn{1}{|p{4.055em}|}{$\epsilon_1$} & Constant for controlling accuracy in  Algorithm 1 &   $\epsilon_2$ &Constant for controlling accuracy in  Algorithm 1 &   $\epsilon_3$ & Constant for controlling accuracy in  Algorithm 1 \\
	\hline
	\multicolumn{1}{|p{4.055em}|}{$\zeta_k$} & The stepsizes corresponding to the related dual variable during iterations  in   \eqref{3C} &    $\xi_{k}$  & The stepsizes corresponding to the related dual variable during iterations  in   \eqref{3C} &   $\theta$ & The stepsizes corresponding to the related dual variable during iterations  in   \eqref{3C}\\
	\hline
\end{tabular}%
\label{tab:addlabel}%
\end{table*}%

The offloading time $t_{k,\text{off}}$ of user $k$ mainly consists of two parts \footnote{In practice, the MEC-integrated BS will provide sufficient transmission power, while the amount of output data from MEC server to user $k$ is usually much less than that of the input data, the time consumed and the transmission energy for delivering the computed results are negligible \cite{Hu2018Wireless}.\label{footnote1}}: the uplink transmission time $t_{k,u}$ from user $k$ to the {\spaceskip=0.3em\relax MEC-integrated BS and the corresponding execution time at MEC server $t_{k,m}$. Therefore, the offloading time $t_{k,\text{off}}$ is given by}
\begin{equation}\label{t_off}
t_{k,\text{off}}=t_{k,u}+t_{k,m}=\frac{\lambda_kR_{k}}{r_{k}}+\frac{\lambda_kR_kc_{k}}{f_{k,m}}.
\end{equation}
	
Due to the parallel computing at users and MEC server, the total latency for user $k$ depends on the larger one between $t_{k,l}$ and $t_{k,\text{off}}$, and can be expressed as
$t_k = \max \{t_{k,l}, t_{k,\text{off}}\}$.
	
\subsection{Energy Consumption}
According to the strategy of computation offloading at user $k$, the total energy consumption comprises two parts\footref{footnote1}: the energy for local computing and for offloading, given in details as follow. 
\subsubsection{Local computing mode}
Given the processor's computing speed $f_k$, the power consumption of the processor is modeled as $\kappa_kf_k^3$ (joule per second), where $\kappa_k$ represents the computation energy efficiency coefficient related to the processor's chip of user $k$ \cite{Zhang2013Energy,Wang2016Mobile,Bi2018Computation}. Taking consideration of \eqref{local time}, the energy consumption at this mode is given by
\begin{equation}
E_{k,l}=\kappa_kf_k^3t_{k,l}=\kappa_kc_k\left(1-\lambda_k\right) R_kf_k^2. 
\end{equation}
\subsubsection{Computation offloading mode}
In this mode, the energy consumption includes the cost of uplink transmission and remote computation for offloaded $\lambda_kR_k$ input bits, which can be obtained as
\begin{equation}\label{e_u,m}
E_{k,\text{off}}=E_{k,u}+E_{k,m}\!=\! \sum_{n\in\mathcal{N}}x_{k,n}p_{k,n}\frac{\lambda_kR_k}{r_k}+\kappa_m\lambda_kc_kR_kf_{k,m}^2,
\end{equation}
where $\kappa_m$ is the computation energy efficiency coefficient related to the processor's chip of MEC server.
	
Therefore, the total energy consumption for user $k$ related with its computation offloading strategy in our system is
\begin{equation}
E_k = E_{k,l}+E_{k,u}+E_{k,m}.
\end{equation}
	
In this paper, we minimize the overall energy consumption of the considered system, which is related to joint optimization on offloading ratio, transmission power, and subcarriers and computing resource allocation. Mathematically, the energy consumption minimization problem can be written as 
\begin{subequations}\label{OP1}
	\begin{align}
	\mathbf{P}:~\min _{\boldsymbol{\lambda},\boldsymbol{p},\boldsymbol{f},\boldsymbol{X}}~&\sum_{k\in\mathcal{K}}E_{k}\\
	\mathrm{s.t.}  ~& 0\leq \lambda_{k}\leq 1,\forall k, \label{OP1-C1}\\
	~&  \max \{t_{k,l}, t_{k,\text{off}}\} \leq T,\forall k,\label{OP1-C2}\\
	~& \sum_{n\in\mathcal{N}} x_{k,n}p_{k,n}\leq p_k^\text{max}, \forall k, \label{OP1-C7}\\
	~&  0 \leq f_{k,m}, \forall k,\label{OP1-C3}\\
	~& \sum_{k\in\mathcal{K}}f_{k,m}\leq F,\label{OP1-C4}\\
	~& \sum_{k\in\mathcal{K}}x_{k, n} \leq 1, \forall n,\label{OP1-C5}\\
	~&  x_{k, n} \in\{0,1\},\forall k, n,\label{OP1-C6}
	\end{align}
\end{subequations}
where $\boldsymbol{\lambda}\triangleq\{\lambda_{k}\}$, $\boldsymbol{p}\triangleq\left\{p_{k,n}\right\}$, $\boldsymbol{f}\triangleq\{f_{k,m}\}$ and $\boldsymbol{X}\triangleq\left\{x_{k,n}\right\}$. The constraints above can be explained as follows: constraint \eqref{OP1-C2} states that the task of user $k$ must be completely executed within a time slot; constraint \eqref{OP1-C7} shows that the total transmission power on the subcarriers of user cannot exceed its maximum transmit budget; constraint \eqref{OP1-C3} and \eqref{OP1-C4} show that MEC server must allocate a positive computing resource to the user associated with it, and the sum of the allocated computing resource cannot exceed the total computation capability of MEC server; constraint \eqref{OP1-C5} and \eqref{OP1-C6} enforce that each subcarrier can only be used by one user to avoid the multi-user interference.

\section{Offloading and Resource Allocation Strategy}\label{3}
In this section, we provide offloading and resource allocation strategy for the considered optimization problem $\mathbf{P}$, which is intractable to deal with due to the coupled variants in both the constraints and the objective function based on our observation. To decouple these variants, we will divide the orignal problem $\mathbf{P}$ into three subprolems: 1) $\mathbf{P_O}$, offloading ratio selction; 2) $\mathbf{P_T}$, transmission power optimization; 3) $\mathbf{P_S}$, subcarriers and computing resource allocation. 

\textbf{Firstly}, with given transmission power $\boldsymbol{p}$, computation capability assignment $\boldsymbol{f}$ and subcarrier allocation strategy $\boldsymbol{X}$, we can obtain the optimal offloading ratio $\boldsymbol\lambda^\star$ at the outer loop\footnote{The outer loop is designed for the three subproblems w.r.t. $\mathbf{P_O}$, $\mathbf{P_T}$ and $\mathbf{P_S}$, and to update the achieved variants iteratively.}. \textbf{Secondly}, with the newly obtained offloading ratio $\boldsymbol{\lambda}^\star$, we can optimize the transmission power $\boldsymbol{p}$ according to the first-order Taylor expansion and the transformation of the sum of fractional functions, while updating dual variable $(\boldsymbol\varphi, \boldsymbol\vartheta)$ and then renewing auxiliary variable $(\boldsymbol{a}, \boldsymbol{b})$ at its inner loop. \textbf{Thirdly}, we can optimize $(\boldsymbol{f}, \boldsymbol{X})$ at one iteration with the newly achieved $(\boldsymbol\lambda^\star,\boldsymbol{p}^\star)$ at its inner loop, and renew auxiliary variable $\boldsymbol{\phi}$. Then, with the newly optimized $(\boldsymbol{f}, \boldsymbol{X})$, we further update the corresponding dual variables $(\boldsymbol\alpha,\boldsymbol\beta,\gamma)$ at the next iteration at its inner loop. \textbf{Finally}, we will iteratively update the derived $(\boldsymbol\lambda,\boldsymbol{p},\boldsymbol{f}, \boldsymbol{X})$ at the outer loop, and the procedures are known as the BCD method \cite{Richt2014Iteration,zhao2017exploiting}. In this section, the joint optimization on offloading ratio, transmission power, and subcarriers and computing resource allocation will be proposed in accordance with the iterative approach based on the BCD method as follows.

\subsection{Offloading selection}\label{3A}
Given the transmission power $\boldsymbol{p}$, the computation capability assignment $\boldsymbol{f}$ and the subcarrier allocation strategy $\boldsymbol{X}$, the optimal offloading ratio $\boldsymbol{\lambda}^\star$ can be obtained by solving the following problem,
\begin{equation}\label{OP1l}
\begin{aligned}
\mathbf{P_O}:~\min _{\boldsymbol{\lambda}}~&\sum_{k\in\mathcal{K}}E_{k}\\
\mathrm{s.t.}  ~&\eqref{OP1-C1}\eqref{OP1-C2},
\end{aligned}
\end{equation}	
which can be decoupled into $K$ subproblems related to each user, given by
\begin{subequations}\label{OP1lA}
	\begin{align}
		\mathbf{P_{O}1}:~\min _{\lambda_{k}}~&E_{k}\\
		\mathrm{s.t.}  ~& 0\leq \lambda_{k}\leq 1,\label{OP1lA-C1}\\
		~&  1-\frac{Tf_{k}}{c_kR_{k}} \leq \lambda_{k} \leq \frac{Tr_kf_{k,m}}{R_kf_{k,m}+r_kR_kc_{k}},\label{OP1lA-C2}
	\end{align}
\end{subequations}
where \eqref{OP1lA-C2} can be derived with the help of \eqref{local time} and \eqref{t_off}. Apparently, $\mathbf{P_{O}1}$ is a convex problem with respect to (w.r.t.) $\lambda_k$, and the optimal offloading ratio $\lambda_k^\star$ at user $k$ can be achieved according to the following theorem.
\begin{theorem}\label{theorem1}
With given $(\boldsymbol{p},\boldsymbol{f},\boldsymbol{X})$, the optimal $\lambda_k^\star$ for $\mathbf{P_{O}1}$ is given by\footnote{The value of $E_{k}$ is independent of $\lambda_{k}$ when $\frac{\partial E_{k}}{\partial \lambda_{k}}=0$, thus we can choose any offloading ratio as the optimal ratio.}		
\begin{equation}\label{OP-lkm}
\lambda_{k}^\star= 
\left\{
\begin{aligned}
	&\max\left\{1-\frac{Tf_{k}}{c_kR_{k}}, 0\right\}, &&\text{if}~\frac{\partial E_{k}}{\partial \lambda_{k}} \geq 0,\\
	&\min \left\{\frac{Tr_kf_{k,m}}{R_kf_{k,m}+r_kR_kc_{k}},1\right\}, &&\text{otherwise}.
\end{aligned}
\right.
\end{equation}
\end{theorem}
\begin{IEEEproof}
It can be obtained resorting to the first-order condition and comparing with the boundaries points provided by \eqref{OP1lA-C1} and \eqref{OP1lA-C2}.
\end{IEEEproof}
\begin{remark}\label{remark1}
Based on the first-order derivative, we can obtain the optimal $\lambda_k^\star$ for user $k$, $\forall k$. When $\frac{\partial E_{k}}{\partial \lambda_{k}} \geq 0$, the $k$-th user is willing to compute its own data locally with the increasing $f_k$, but tends to offload more data to the MEC-integrated BS when it has very limited computation capability, i.e., low CPU frequency at local. When $\frac{\partial E_{k}}{\partial \lambda_{k}} < 0$, the $k$-th user prefers offloading to the MEC-integrated BS if MEC server assigns a growing number of computation resource, $f_{k,m}$ to the user, or the user has a gradual increasing transmission rate $r_k$, and vice versa.
\end{remark}

To get more insight of the above theorem, we can obtain the following corollary based on some asymptotic assumptions, and further get some meaningful conclusions.  
\begin{corollary}
When the CPU of user $k$ is very powerful such that $f_k\rightarrow+\infty$, or MEC server assigns a large enough computation capability for user $k$ such that $f_{k,m}\rightarrow+\infty$, the optimal $\lambda_k^\star$ that minimizes the energy consumption in $\mathbf{P_{O}1}$ is given as
\begin{equation}
\lambda_{k}^\star= 
\left\{
\begin{aligned}
&1-\frac{Tf_{k}}{c_kR_{k}}, &&\text{if}~f_{k,m}\rightarrow+\infty,\\
&\frac{Tr_kf_{k,m}}{R_kf_{k,m}+r_kR_kc_{k}}, &&\text{if}~f_k\rightarrow+\infty.
\end{aligned}
\right.
\end{equation}
\end{corollary}
\begin{IEEEproof}
The first-order condition of $E_k$ is given by
\begin{equation}
\frac{\partial E_{k}}{\partial \lambda_{k}}=-\kappa_kc_kR_kf_k^2+\sum_{n\in\mathcal{N}_k}p_{k,n}\frac{R_k}{r_k}+\kappa_mc_kR_kf_{k,m}^2.\nonumber
\end{equation}
When the CPU frequency of the $k$-th user approaches to infinity, i.e., $f_k\rightarrow+\infty$, it is straightforward that $\frac{\partial E_{k}}{\partial \lambda_{k}}\ll 0$, and thus we have $\lambda_{k}^\star=\frac{Tr_kf_{k,m}}{R_kf_{k,m}+r_kR_kc_{k}}$ referring to \eqref{OP-lkm}. Similarly, we can obtain the optimal $\lambda_k^\star$ when $f_{k,m}\rightarrow+\infty$.
\end{IEEEproof}

\begin{remark}\label{remark2}
From Theorem \ref{theorem1}, we observe that user $k$ will compute data locally by offloading a part of data, i.e., $\left(1-\frac{Tf_{k}}{c_kR_{k}}\right)$ to MEC server as its CPU frequency $f_k$ is growing when $\frac{\partial E_{k}}{\partial \lambda_{k}}\geq 0$. Notice that when the increasing $f_k$ exceeds a certain value and thus makes $\frac{\partial E_{k}}{\partial \lambda_{k}}< 0$, user $k$ prefers to offload its data based on $\left(\frac{Tr_kf_{k,m}}{R_kf_{k,m}+r_kR_kc_{k}}\right)$, rather than computes its whole data locally even though $f_k\rightarrow+\infty$. The reason is that MEC server has more efficient computation energy coefficient than the users, i.e., $\kappa_m<\kappa_k,\forall k$, and thus users are inclined to offload partial data to reduce the energy consumption even if they have sufficiently large CPU frequencies. Moreover, when the growing $f_{k,m}$ is over the threshold which makes $\frac{\partial E_{k}}{\partial \lambda_{k}}<0$ into $\frac{\partial E_{k}}{\partial \lambda_{k}}\geq0$ according to \eqref{OP-lkm}, user $k$ will choose $\left(1-\frac{Tf_{k}}{c_kR_{k}}\right)$ as its offloading ratio. That is because users will have additional energy consumption and communication delay while uploading data to the MEC-integrated BS even though MEC server has more efficient computation. Taking $E_{k,l}$ and $E_{k,\text{off}}$ into consideration, it is more beneficial for users to compute partial data locally to reduce the energy consumption while satisfying the QoE requirement of time-sensitive computation tasks at users, in despite of the assigned large enough computation capability.
\end{remark}
	
	%%%%%%%%%%%%%%%%%%%%%%%%%%%%%算法流程
\subsection{Transmission power optimization}\label{3B}
In this subsection, with the newly obtained offloading ratio $\boldsymbol{\lambda}^\star$, the optimal transmission power $\boldsymbol{p}^\star$ can be obtained by solving the following problem with the given computation capability assignment $\boldsymbol{f}$ and subcarrier allocation strategy $\boldsymbol{X}$, 
\begin{subequations}\label{OP2-P1}
	\begin{align}
	\mathbf{P_T}:~\min_{\boldsymbol{p}}~&\sum_{k\in\mathcal{K}}E_{k,u}\label{OP2-OF}\\
	\mathrm{s.t.} ~& \sum_{n\in\mathcal{N}_k} p_{k,n}\leq p_k^\text{max},\forall k,\label{OP2-C1}\\
	~& t_{k,u}+t_{k,m}\leq T, \forall k,\label{OP2-CT}
	\end{align}
\end{subequations}
where $E_{k,u}\!=\!\sum_{n\in\mathcal{N}_k}p_{k,n}\frac{\lambda_kR_{k}}{{\sum_{n\in\mathcal{N}_k}B\log_{2}\left(1+p_{k,n}\tilde{g}_{k, n}\right)}}$, $\mathcal{N}_k$ represents the set of subcarriers allocated to user $k$, and $N_k$ represents the size of the related set. Based on the observation of $\mathbf{P_T}$, it is clear that $\mathbf{P_T}$ is non-convex since the objective function involves sum-of-ratio minimization, and constraint \eqref{OP2-CT} further makes the optimization problem even more difficult to tackle.

To address the above issues, we first deal with the non-convex constraint \eqref{OP2-CT}. Based on \eqref{r_k} and \eqref{t_off}, the constraint for the QoE requirement of time-sensitive computation tasks \eqref{OP2-CT} can be recast as  
\begin{equation}\label{OP-CTU}
\begin{aligned}
\sum_{n\in\mathcal{N}_{k}}\log_{2}(1+p_{k,n}\tilde{g}_{k,n}) \geq \frac{\lambda_{k}R_{k}}{BT_{k,u}}, \forall k, 
\end{aligned}    
\end{equation}  
where $T_{k,u}\triangleq T-\frac{\lambda_{k}R_{k}c_{k}}{f_{k,m}}$. However, the transformed contraint \eqref{OP-CTU} is still non-convex since the summation of concave functions is greater than or equal to the constant. In order to deal with the non-convexity of \eqref{OP-CTU}, we obtain the following inequalities by applying the first-order Taylor expansion at the given points $p_{k,n}^r,\forall k,n$ in the $r$-th iteration,
\begin{align}\label{OP-Taylor}
&\sum_{n\in\mathcal{N}_{k}}\log_{2}\left(1+p_{k,n}\tilde{g}_{k,n}\right) \geq\\
&\sum_{n\in\mathcal{N}_{k}}\left\{\log_2\left(1+p_{k,n}^r\tilde{g}_{k,n}\right)+\frac{\tilde{g}_{k,n}\left(p_{k,n}-p_{k,n}^r\right)}{\left(1+p_{k,n}^r\tilde{g}_{k,n}\right)\ln 2}\right\}, \forall k,\nonumber
\end{align} 
where the right-hand side of \eqref{OP-Taylor} is now a linear function w.r.t. $p_{k,n}$. With some mathematic manipulations, constraint \eqref{OP-CTU} can be recast as
\begin{equation}\label{Talyor-C1}
\sum_{n\in\mathcal{N}_{k}}\frac{\tilde{g}_{k,n}p_{k,n}}{\left(1+p_{k,n}^r\tilde{g}_{k,n}\right)\ln 2} \geq o_k, \forall k,
\end{equation}
where we define $o_k\triangleq\frac{\lambda_{k}R_{k}}{BT_{k,u}}+ \sum_{n\in\mathcal{N}_{k}}\left\{\frac{\tilde{g}_{k,n}p_{k,n}^r}{\left(1+p_{k,n}^r\tilde{g}_{k,n}\right)\ln 2}\right\}-\log_2\left(1+p_{k,n}^r\tilde{g}_{k,n}\right)$.

Next, we start to tackle the non-convex objective function in \eqref{OP2-P1}, which is in the form of sum of ratios. For notational brevity, we denote 
\begin{eqnarray}
	&&d_{k}(\boldsymbol{p}_k)=\sum_{n\in\mathcal{N}_{k}}p_{k,n}\lambda_kR_{k}, \label{OP-PD} \\
	&&h_{k}(\boldsymbol{p}_k)=B\sum_{n\in\mathcal{N}_k}\log_2(1+p_{k,n}\tilde{g}_{k,n}),\label{OP-PH}
\end{eqnarray}
where $\boldsymbol{p}_k=\{p_{k,n}\},\forall n\in\mathcal{N}_k$. By introducing auxiliary variables $\boldsymbol{a}$, where $\boldsymbol{a}\triangleq\{a_k\}$, the original problem with some simple transformations can be rewritten as
\begin{subequations}\label{OP2-P2}
	\begin{align}
	\mathbf{P_T1}:~\min _{\boldsymbol{p},\boldsymbol{a}}~&\sum_{k\in\mathcal{K}}a_{k}\\
	\mathrm{s.t.}  
	~& \eqref{OP2-C1}\eqref{Talyor-C1},\nonumber\\
	~& d_{k}(\boldsymbol{p}_k)\leq a_kh_{k}(\boldsymbol{p}_k),\forall k, \label{OP2-C2}
	\end{align}
\end{subequations}
which is still difficult to solve due to the coupled variables in the constraints. Therefore, we employ the equivalent parametric convex programming described in the following lemma to cope with $\mathbf{P_T1}$.
\begin{lemma}\cite[Lemma 2.1]{Yjong2012}\label{op-lemma}
	For $\forall k, n$, if $\left(\{p_{k,n}^\star\},\{a_k^\star\}\right)$ is the the optimal solution of $\mathbf{P_T1}$, there must exist $\{b_k^\star\}$ such that $\{p_{k,n}^\star\}$ satisfies the Karush-Kuhn-Tucker (KKT) condition of the following problem for $b_{k}=b_{k}^\star$ and $a_{k}=a_{k}^\star$,
	\begin{subequations}\label{OP2-P3}
		\begin{align}
		\mathbf{P_{T}2}:~\min_{\boldsymbol{p}}~&\sum_{k\in\mathcal{K}}b_{k}\left[d_{k}(\boldsymbol{p}_k)-a_kh_{k}(\boldsymbol{p}_k)\right]\\
		\mathrm{s.t.}  
		~& \eqref{OP2-C1}\eqref{Talyor-C1}.\nonumber
		\end{align}
	\end{subequations}
	
	Furthermore, $\{p_{k,n}^\star\}$ satisfies the following equations for $b_{k}=b_{k}^\star$ and $a_{k}=a_{k}^\star$:\\
	\begin{align}\label{OP2-b}
	b_{k} = \frac{1}{h_{k}(\boldsymbol{p}_k)},~a_{k} = \frac{d_{k}(\boldsymbol{p}_k)}{h_{k}(\boldsymbol{p}_k)},~\forall k,
	\end{align}
\end{lemma}
where $\{b_k\}$ is the non-negative multiplier of constraint \eqref{OP2-C2}, and we define $\boldsymbol{b}\triangleq\{b_k\}$ for simplicity.
\begin{IEEEproof}
It can be proved by taking the derivative of the Lagrange function of $\mathbf{P_T1}$, the details of which can be referred to \cite{Yjong2012}, and thus is omitted for brevity. 	
\end{IEEEproof}

\textit{Lemma} \ref{op-lemma} implies that the optimal solution of $\mathbf{P_T1}$ can be obtained by solving the equations of \eqref{OP2-b} among the solution of $\mathbf{P_T2}$. The Lagrangian of $\mathbf{P_T2}$ is
\begin{align}\label{OP-LF}
&\mathcal{L}(\boldsymbol{p}, \boldsymbol{\varphi}, \boldsymbol{\vartheta})\nonumber\\
&=\sum_{k\in\mathcal{K}}b_{k}\left[d_{k}(\boldsymbol{p}_k)\!-\!a_kh_{k}(\boldsymbol{p}_k)\right]\!+\!\sum_{k\in\mathcal{K}}\varphi_{k}\left(\sum_{n\in\mathcal{N}_k}p_{k,n}\!-\!p_k^\text{max}\right)\nonumber\\
&+\sum_{k\in\mathcal{K}}\vartheta_{k}\left[o_k-\sum_{n\in\mathcal{N}_{k}}\frac{\tilde{g}_{k,n}p_{k,n}}{\left(1+p_{k,n}^r\tilde{g}_{k,n}\right)\ln 2}\right],
\end{align}
where $\boldsymbol{\varphi}\triangleq\{\varphi_k\},~\boldsymbol{\vartheta}\triangleq\{\vartheta_k\},\forall k$ are the non-negative Lagrange multipliers for the corresponding constraints. It can be readily proved that $\mathbf{P_T2}$ is convex for given $b_k$ and $a_k,\forall k$, and satisfies Slater’s condition. Thus, strong duality holds between the primal and dual problems, which means solving $\mathbf{P_T2}$ is equivalent to solving the dual problem. Define $\mathcal{P}$ as all sets of possible $\boldsymbol{p}$ that satisfy constraint \eqref{OP2-C1}. Notice that the Lagrange dual function is then defined as
\begin{equation}\label{OP-DF1}
g(\boldsymbol{\varphi}, \boldsymbol{\vartheta})=\min_{\boldsymbol{p}\in \mathcal{P}}\mathcal{L}(\boldsymbol{p}, \boldsymbol{\varphi}, \boldsymbol{\vartheta}). 
\end{equation}
Moreover, the Lagrange dual problem becomes
\begin{equation}\label{OP2-DP}
\begin{aligned}
\max_{\boldsymbol{\varphi}, \boldsymbol{\vartheta}} ~&g(\boldsymbol{\varphi}, \boldsymbol{\vartheta})\\
\mathrm{s.t.} ~&\boldsymbol{\varphi}\succeq \boldsymbol{0},~\boldsymbol{\vartheta}\succeq \boldsymbol{0}.
\end{aligned}    
\end{equation}
In the following, we obtain the optimal transmission power $\boldsymbol{p}$ for given auxiliary variables ($\boldsymbol{b},\boldsymbol{a}$) and Lagrange multipliers ($\boldsymbol{\varphi}, \boldsymbol{\vartheta}$) at first, then the Lagrange multipliers are updated via gradient descent method. At last, the auxiliary variables are updated based on iterative method as well.
\subsubsection{Update $p_{k,n}$}\label{OP2-P}
To minimize the Lagrange dual function in \eqref{OP-DF1}, $g(\boldsymbol\varphi, \boldsymbol\vartheta)$ can be decomposed into $K$ independent {\spaceskip=0.2em\relax
subproblems. To be specific, the subproblem for user $k$ is given by}
\begin{equation}\label{OP-DF2}
g(\varphi_k,\vartheta_k)=\min _{{\boldsymbol{p}_k}\in \mathcal{P}}\mathcal{L}(\boldsymbol{p}_k,\varphi_{k},\vartheta_{k}),
\end{equation}
where 
\begin{align}\label{LG-UE}
\mathcal{L}(\boldsymbol{p}_k,\varphi_{k},\vartheta_{k})=&b_{k}\left[d_{k}(\boldsymbol{p}_k)\!-\!a_kh_{k}(\boldsymbol{p}_k)\right]\!+\!\varphi_{k}\sum_{n\in\mathcal{N}_k}p_{k,n}\nonumber\\
&-\vartheta_{k}\sum_{n\in\mathcal{N}_{k}}\frac{\tilde{g}_{k,n}p_{k,n}}{\left(1+p_{k,n}^r\tilde{g}_{k,n}\right)\ln 2}.
\end{align}
On the observation of \eqref{LG-UE}, it can be further decomposed into $|N_k|$ subproblems w.r.t. the subcarriers allocated to user $k$, given as
\begin{equation}\label{OP-DF2-SC}
g_n(\varphi_k,\vartheta_k)=\min _{{p_{k,n}}\in \mathcal{P}}\mathcal{L}(p_{k,n},\varphi_{k},\vartheta_{k}),
\end{equation}
where 
\begin{align}\label{LG-UE-SC}
\mathcal{L}(p_{k,n},\varphi_{k},\vartheta_{k})=&b_{k}\left[p_{k,n}\lambda_kR_{k}-a_kB\log_2\left(1+p_{k,n}\tilde{g}_{k,n}\right)\right]\nonumber\\
&+\varphi_{k}p_{k,n}-\frac{\vartheta_{k}\tilde{g}_{k,n}p_{k,n}}{\left(1+p_{k,n}^r\tilde{g}_{k,n}\right)\ln 2},
\end{align}
combined with \eqref{OP-PD} and \eqref{OP-PH}.
\begin{theorem}\label{th1}
	The optimal transmission power for the $k$-th user at the $n$-th subcarrier is $p_{k,n}^\star$, where $n\in\mathcal{N}_k$, given by
	\begin{equation}\label{pkn_op}
	p_{k,n}^\star=\left[\frac{a_kb_kB}{\left(b_k\lambda_kR_k+\varphi_k\right)\ln 2\!-\!\frac{\vartheta_k\tilde{g}_{k,n}}{1+p_{k,n}^r\tilde{g}_{k,n}}}\!-\!\frac{1}{\tilde{g}_{k,n}}\right]^+.
	\end{equation}
\end{theorem}
\begin{IEEEproof}
The proof is presented in Appendix A.
\end{IEEEproof}
\begin{remark}
To minimize the energy consumption of offloading from users to MEC, $\sum_{k\in\mathcal{K}}E_{k,u}$, user $k$ will distribute more power to its allocated subcarrier with a higher channel gain $\tilde{g}_{k,n}$, and the assigned power is proportional to the bandwidth of subcarrier, $B$. These can be seen from the optimal expression of $p_{k,n}^\star$. On the other hand, user $k$ is inclined to reduce its transmission power when it plans to offload more data to the MEC-integrated BS, in order to reduce the energy consumption of offloading, which can be deduced from the formulation of $E_{k,u}$ in \eqref{OP2-OF}.
\end{remark}

Interestingly, an asymptotic solution can be obtained in strong channel gain scenario, presented by the following corollary.
\begin{corollary}\label{coro1}
When the channel gain for user $k$ at the $n$-th subcarrier is very strong such that  $\tilde{g}_{k,n}\rightarrow +\infty$, the optimal transmission power $p_{k,n}^\star$ is given by
\begin{equation}
	p_{k,n}^\star=\left[\frac{a_kb_kB}{\left(b_k\lambda_kR_k+\varphi_k\right)\ln 2-\frac{\vartheta_k}{p_{k,n}^r}}\right]^+,
\end{equation} 
where $p_{k,n}^\star$ is irrelevant to the channel gain.
\end{corollary}
\begin{IEEEproof}
It can be directly deduced from \eqref{pkn_op}, and thus we omit it.
\end{IEEEproof}

\subsubsection{Lagrange Multipliers Update}\label{3B-UL}
With the achieved $\boldsymbol{p}^\star$, we start to update the Lagrange multiplier $(\boldsymbol\varphi, \boldsymbol\vartheta)$. With known $\boldsymbol{p}^\star$, the Lagrange dual problem \eqref{OP2-DP} is always convex, which can be decomposed into $K$ independent subproblems w.r.t. each user, and the one related to user $k$ is given as
\begin{equation}\label{OP2-DP-sub}
\begin{aligned}
\max_{\varphi_{k}, \vartheta_{k}} ~&g(\varphi_{k}, \vartheta_{k})\\
\mathrm{s.t.} ~&\varphi_k\geq 0,~\vartheta_k\geq 0,
\end{aligned}    
\end{equation}
which is an affine function w.r.t dual variables. Thus, we can apply the simple gradient method for the variable update, and the dual variables $(\varphi_{k}, \vartheta_{k})$ can be updated according to the following formulations. For example, $(\varphi_{k}^{j+1}, \vartheta_{k}^{j+1})$ can be obtained at the $(j+1)$-th iteration, respectively, as
\begin{equation}\label{l1}
\varphi_{k}^{j+1}=\left[\varphi_{k}^{j}-\iota_{k}\left(\sum_{n\in\mathcal{N}_k}p_{k,n}\!-\!p_k^\text{max}\right)\right]^{+},
\end{equation}
\begin{equation}\label{l2}
\vartheta_{k}^{j+1}\!=\!\left\{\!\vartheta_{k}^{j}\!-\!\nu_{k}\left[o_k\!-\!\!\sum_{n\in\mathcal{N}_{k}}\frac{\tilde{g}_{k,n}p_{k,n}}{\left(1+p_{k,n}^r\tilde{g}_{k,n}\right)\ln 2}\right]\!\right\}^{+},
\end{equation}
where $\iota_{k}$ and $\nu_{k}$ are the stepsizes corresponding to the related dual variable during iterations. Notice that all the Lagrange variables must be non-negative, the details of which are given in TABLE \uppercase\expandafter{\romannumeral2}. 
\subsubsection{Auxiliary Variables Update}\label{3B-UA}
Then, we can update the auxiliary variables $(\boldsymbol{a}, \boldsymbol{b})$ in this subsection. According to \textit{Lemma} \ref{op-lemma}, the optimal $\boldsymbol{p}_k^{\star}$ also satisfies the following conditions: 
\begin{align}\label{OP2-UA2}
~&a_kh_{k}(\boldsymbol{p}_k^{\star})-d_{k}(\boldsymbol{p}_k^{\star})=0,~\forall k,\\
~&b_kh_{k}(\boldsymbol{p}_k^{\star})-1=0,~\forall k.
\end{align}

Similarly, referring to \cite{Yjong2012}, we define functions for notational brevity. Specifically, let $W_{n}(a_{n})\!=\!a_{n}h_{k}(\boldsymbol{p}_k^{\star})\!-\!d_{k}(\boldsymbol{p}_k^{\star})$ and $W_{n+K}(b_n)\!=\!b_{n}h_{k}(\boldsymbol{p}_k^{\star})\!-\!1$, $n\in\{1,2,...,K\}$. The optimal solutions of $(b_{k}^\star,a_{k}^\star)$ can be obtained by solving $\boldsymbol{W}(b_{k},a_{k})\triangleq\left[W_{1}, W_{2},...W_{2K}\right]=0$. 
Then, we can apply iterative methods to update auxiliary variables 
\begin{align}
~&b_{k}^{i+1}=(1-\tau^i)b_{k}^i+\frac{\tau^i}{h_{k}(\boldsymbol{p}_k^{\star})}\label{OP2-UAb},\forall k,\\
~&a_{k}^{i+1}=(1-\tau^i)a_{k}^i+\tau^i\frac{d_{k}(\boldsymbol{p}_k^{\star})}{h_{k}^{\star}(\boldsymbol{p}_k^{\star})},\forall k,\label{OP2-UAa}
\end{align}
where $\tau^i$ is the largest $\tau$ that satisfies 
\begin{equation}
\|\boldsymbol{W}\left(b_{k}^i+\tau^{l}\boldsymbol{\mathrm{q}}_{K+1:2K}^{i},a_{k}^i\!+\!\tau^{l}\boldsymbol{\mathrm{q}}_{1:K}^{i}\right)\|\!\leq\!(1-z\tau^{l})\|W(b_{k}^i,a_{k}^i)\|,
\end{equation}
where $\boldsymbol{q}$ is the Jacobian matrix of $\boldsymbol{W},~l\in\{1,2,\cdots\}$, $\tau_{l}\in(0,1)$ and $z\in (0,1).$ Note that when $\tau^i=1$, it becomes the standard Newton method, and the detailed process of achieving $\boldsymbol{p}^{\star}$ is summarized in Algorithm 1.
\begin{algorithm}[h]
	\caption{Transmission Power Algorithm}
	{\bf Input:} %算法的输入， \hspace*{0.02in}用来控制位置，同时利用 \\ 进行换行
	Given offloading ratio $\boldsymbol\lambda^\star$, computation
	capabilities assignment $\boldsymbol{f}^\star$, and subcarrier allocation
	strategy $\boldsymbol{X}^\star$. \\
	{\bf Output:} %算法的结果输出
	The optimal $\boldsymbol{p^{\star}}$
	\begin{algorithmic}[1]
		\State {\bf Initialization:} give iteration $r\!=\!0$, and the algorithm accuracy indicators $(\epsilon_1,\epsilon_2)$, where $\epsilon_1$ and $\epsilon_2$ are a very small constant for controlling accuracy.
		\Repeat (from 2 to 18)
			\State With given local points $p_{k,n}^{r}$.
	   	 	\State {\bf Initialization:} auxiliary variables $\boldsymbol{b}$ and $\boldsymbol{a}$. 
	    	\Repeat (from 5 to 14)
				\State {\bf Initialization:} Lagrange variables $\boldsymbol\varphi$
and $\boldsymbol\vartheta$ 
				\Repeat(from 7 to 12)
					\Repeat (from 8 to 11)
		 			\State Obtain transmission power $p_{k,n}$.
		 			\Until Lagrangian function converges.
				    \State Update Lagrange variables based on gradient descent
method in \eqref{l1} and \eqref{l2}.
				\Until {$\boldsymbol\varphi$ and $\boldsymbol\vartheta$ converge.}
			\State update auxiliary variables from \eqref{OP2-UAb} and \eqref{OP2-UAa}
		\Until {$\| \boldsymbol{\mathrm{W}}(b_{k},a_{k}) \|\!\leq\!\epsilon_1 $}
		\State Output the optimal $p_{k,n}^{\star}$
		\State Update local point $p_{k,n}^{r+1}=p_{k,n}^{\star}$
		\State $r=r+1$
		\Until $|p_{k,n}^{r+1}-p_{k,n}^{r}|\!\leq\!\epsilon_2$
	\end{algorithmic}	
\end{algorithm}

\subsection{Subcarriers and Computing Resource Allocation}\label{3C}
With the newly achieved offloading ratio $\lambda^\star$ and transmission power $\boldsymbol{p}^\star$, the subcarriers and computing resource allocation will be designed to assign computation capability of MEC server, and allocate subcarriers for each user, to further reduce the energy consumption. In this subsection, we aim to minimize the energy consumption for computation offloading mode via the following optimization problem, the objective function of which is $\sum_{k\in\mathcal{K}}E_{k,\text{off}}$ indeed,
\begin{subequations}\label{OP-C}
	\begin{align}
	\mathbf{P_S}:~\min _{\boldsymbol{f},\boldsymbol{X}}~&\sum_{k\in\mathcal{K}}\sum_{n\in\mathcal{N}}x_{k,n}p_{k.n}\frac{\lambda_kR_k}{r_k}\!+\!\kappa_m\lambda_kc_kR_kf_{k,m}^2\\
	\mathrm{s.t.}
	~& \eqref{OP1-C7}-\eqref{OP1-C6}, \eqref{OP2-CT}.\nonumber
	\end{align}
\end{subequations}
The main challenge in solving $\mathbf{P_S}$ is that the considered optimization problem is a mixed integer programming and thus is NP-hard and non-convex, finding the optimal solution is generally prohibitively due to the computation complexity. Fortunately, the duality gap becomes zero in multi-carrier systems as the number of subcarriers goes to infinity according to the time-sharing condition \cite{Seong2006Optimal,Wei2006Dual}. Thus, the optimal solution for a non-convex resource allocation problem in multi-carrier system can be achieved in the dual domain. 

However, we cannot transform the primal domain of $\mathbf{P_S}$ into the dual domain directly since $r_k$ is in the denominator and its form in \eqref{r_k} makes the problem more intractable. Therefore, a new non-negative auxiliary variable $\boldsymbol{\phi}\triangleq\{\phi_{k}\}$ will be introduced to transform $\mathbf{P_S}$ into the following problem,
\begin{subequations}\label{OP3}
	\begin{align}
	\mathbf{P_S1}:~\min _{\boldsymbol{f},\boldsymbol{X},\boldsymbol{\phi}}~&\sum_{k\in\mathcal{K}}\sum_{n\in\mathcal{N}}x_{k,n}p_{k,n}\frac{\lambda_kR_k}{\phi_k}+\kappa_m\lambda_kc_kR_kf_{k,m}^2\\
	\mathrm{s.t.}
	~&\eqref{OP1-C7}-\eqref{OP1-C6}, \eqref{OP2-CT},\nonumber\\  
	~& 0\leq\phi_{k}\leq r_{k}, \forall k, \label{OPC-C2}
	\end{align}
\end{subequations}
where we can update $\boldsymbol{\phi}$ to help minimize the energy consumption. The Lagrangian for the above problem with the achieved $\boldsymbol{\lambda}^\star$ and $\boldsymbol{p}^\star$ can be written as 
\begin{align}\label{OP3-L}
&\!\mathcal{L}(\boldsymbol{f},\boldsymbol{X},\boldsymbol\phi,\boldsymbol\alpha,\boldsymbol\beta,\boldsymbol\delta,\gamma)\nonumber\\
&=\sum_{k\in\mathcal{K}}\sum_{n\in\mathcal{N}}x_{k,n}p_{k,n}\frac{\lambda_kR_k}{\phi_k}+\kappa_m\lambda_kc_kR_kf_{k,m}^2\\
&~~+\sum_{k\in\mathcal{K}}\alpha_{k}\left(\frac{\lambda_kR_{k}}{\phi_{k}}+\frac{\lambda_kR_kc_{k}}{f_{k,m}}-T\right)+\sum_{k\in\mathcal{K}}\delta_{k}(\phi_k-r_{k})\nonumber\\
&~~+\sum_{k\in\mathcal{K}}\beta_{k}\!\left(\sum_{n\in\mathcal{N}}x_{k,n}p_{k,n}-p_{k}^{\text{max}}\right)+\gamma\!\left(\sum_{k\in\mathcal{K}}f_{k,m}-F\right)\nonumber,
\end{align}
where 
$\boldsymbol\alpha\triangleq\{\alpha_k\}$, $\boldsymbol\beta\triangleq\{\beta_k\}$, $\boldsymbol\delta\triangleq\{\delta_k\}$ and $\gamma$ are the non-negative Lagrange multipliers corresponding to the related contraints. Define $\mathcal{F}$ as all sets of possible $\boldsymbol{f}$ that satisfy constraint \eqref{OP1-C3}, $\mathcal{X}$ as all sets of possible $\boldsymbol{X}$ that satisfy constraints \eqref{OP1-C5}, \eqref{OP1-C6} and \eqref{OP2-CT}, and $\mathcal{Q}$ as all sets of possible $\boldsymbol{\phi}$ that satisfy constraint \eqref{OPC-C2}. The Lagrange dual function is then defined as
\begin{equation}\label{OP3-DF}
\begin{aligned}
\mathbf{g}(\boldsymbol\alpha,\boldsymbol\beta,\boldsymbol\delta,\gamma) = \min_{\boldsymbol{f}\in\mathcal{F},\boldsymbol{X}\in \mathcal{X},\boldsymbol{\phi}\in\mathcal{Q}}\mathcal{L}(\boldsymbol{f},\boldsymbol{X},\boldsymbol\phi,\boldsymbol\alpha,\boldsymbol\beta,\boldsymbol\delta,\gamma).
\end{aligned}    
\end{equation}
Furthermore, the Lagrange dual problem is given by
\begin{equation}\label{OP3-DP}
\begin{aligned}
\max ~&\mathbf{g}(\boldsymbol\alpha,\boldsymbol\beta,\boldsymbol\delta,\gamma)\\
\mathrm{s.t.} ~&\boldsymbol{\alpha}\succeq \boldsymbol{0},~\boldsymbol{\beta}\succeq \boldsymbol{0},~\boldsymbol{\delta}\succeq \boldsymbol{0},
~\gamma \geq 0.
\end{aligned}    
\end{equation}

With the achieved offloading ratio and transmission power, the following steps for updating are adopted to obtain the optimal solutions for computing resource and subcarrier allocation. 
%%%%%%%%%%%%%%%%%%%%%%%%%%%%%%%%%%%%%%%%%%
\subsubsection{Computation Capability Assignment}
Employing the KKT conditions, the following condition is both necessary and sufficient for computation capability assignment's optimality:
%按照正确的求导来做
\begin{align}\label{OP3-fkm}
&\frac{\partial\mathcal{L}(\boldsymbol{f},\boldsymbol{X},\boldsymbol\phi,\boldsymbol\alpha,\boldsymbol\beta,\boldsymbol\delta,\gamma)}{\partial{f_{k,m}}}\nonumber\\
&~~= 2f_{k,m}\kappa_m\lambda_kR_kc_{k}-\frac{\alpha_kc_kR_{k}\lambda_{k}}{f_{k,m}^2}+\gamma=0.
\end{align}

However, it is difficult to find a closed-form expression for the optimal solution, $f_{k,m}^\star$. Fortunately, we can resort to the following proposition to obtain $f_{k,m}^\star$.
\begin{proposition}\label{pro0}
$\mathcal{L}$ is a convex function of $f_{k,m}$.
\end{proposition}
\begin{IEEEproof}
The proof is given in Appendix C.
\end{IEEEproof}

Since $\mathcal{L}$ is a convex function of $f_{k,m}$, and $\frac{\partial\mathcal{L}}{\partial{f_{k,m}}}$ increases monotonically along with $f_{k,m}$, we can adopt the bisection method to obtain $f_{k,m}^\star$ within $0\leq f_{k,m}\leq F$. The detailed process of achieving $f_{k,m}^\star$ is summarized in Algorithm 2.
\begin{algorithm}[h]  
	\caption{Proposed Binary Search Algorithm}  
	{\bf Input:} given offloading ratio $\boldsymbol\lambda^\star$, transmission power $\boldsymbol{p}^\star$, and the algorithm accuracy indicator $\epsilon_3$.
	\begin{algorithmic}[1]
		\For {$k\in\mathcal{K}$}
		\State {\bf Initialize:} $f_{k,m}^\text{UB} = F$ {\bf{and}} $f_{k,m}^\text{LB}=0$.
		\Repeat
		\State Set $f_{k,m}$ = $\frac{1}{2}(f_{k,m}^\text{UB}+f_{k,m}^\text{LB})$.
		\State Compute $\frac{\partial{L}}{\partial{f_{k,m}}}$ according to \eqref{OP3-fkm}.
		\If{$\frac{\partial{\mathcal{L}}}{\partial{f_{k,m}}}>0$}
		\State set $f_{k,m}^\text{UB} = f_{k,m}$;
		\Else
		\State set $f_{k,m}^\text{LB} = f_{k,m}$.
		\EndIf 
		\Until {$\left|\frac{\partial \mathcal{L}}{\partial f_{k,m}}\right|\leq\epsilon_3$.}
		\State Obtain the optimal $f_{k,m}^\star$.
		\EndFor
	\end{algorithmic}  
\end{algorithm}

\subsubsection{Subcarrier Allocation Strategy}\label{SAS}
%%%%%%%%%%%%%%%%%%%%%求解最优的X
When the optimal computing capability assignment is achieved, the optimal subcarrier allocation can be obtained through the following procedures. With some mathematic manipulations, we can rewrite \eqref{OP3-L} as
\begin{align}\label{OP3-L1}
&\mathcal{L}(\boldsymbol{X},\boldsymbol\phi,\boldsymbol\alpha,\boldsymbol\beta,\boldsymbol\delta,\gamma)\nonumber\\
&=\sum_{k\in\mathcal{K}}\!\sum_{n\in\mathcal{N}}x_{k,n}\mathcal{L}_{k,n}\!+\!\sum_{k\in\mathcal{K}}\left(\gamma f_{k,m}+\delta_{k}\phi_k-\beta_{k}p_{k}^{\text{max}}\right)\\
&~~+\!\sum_{k\in\mathcal{K}}\alpha_{k}\!\left(\frac{\lambda_kR_{k}}{\phi_{k}}\!+\!\frac{\lambda_kR_kc_{k}}{f_{k,m}}\!-\!T\right)\!\!+\!\kappa_m\lambda_kc_kR_kf_{k,m}^2\!-\!\gamma F\nonumber,\nonumber
\end{align}
where 
\begin{equation}\label{OP3-W}
\mathcal{L}_{k,n}=\frac{p_{k,n}\lambda_kR_k}{\phi_k}\!+\!p_{k,n}\beta_k\!-\!B\delta_k\log_2\left(1\!+\!p_{k,n}\tilde{g}_{k,n}\right).  
\end{equation}

On the observation of \eqref{OP3-L1}, we further suppose that the $n$-th subcarrier is assigned to user $k$, and thus have the Lagrangian for user $k$,
\begin{equation}\label{OP3-L2}
\begin{aligned}
\mathcal{L}_k=\sum_{n\in\mathcal{N}_k}\mathcal{L}_{k,n}+\omega_{k}-\gamma F,
\end{aligned}    
\end{equation}
where 
\begin{equation}\label{OP3-Ln1}
\begin{aligned}
\omega_{k}=&\gamma f_{k,m}+\delta_{k}\phi_k-\beta_{k}p_{k}^{\text{max}}+\kappa_m\lambda_kc_kR_kf_{k,m}^2\\
&+\alpha_{k}\left(\frac{\lambda_kR_{k}}{\phi_{k}}+\frac{\lambda_kR_kc_{k}}{f_{k,m}}-T\right).
\end{aligned}
\end{equation}
Thus, the subproblem is given by
\begin{equation}\label{OP3-SUB}
\begin{aligned}
\min_{\boldsymbol{x}_{n}\in\mathcal{X}}\mathcal{L}_{k,n}(\boldsymbol{x}_{n},\boldsymbol{\phi},\boldsymbol{\beta},\boldsymbol{\delta})
\end{aligned}    
\end{equation}
where $\boldsymbol x_{n}\triangleq\left\{x_{k, n}\right\},\forall k$, and this problem can be solved independently.
To minimize each $\mathcal{L}_{k,n}$, the optimal $\boldsymbol{x}_n$
can be obtained as
\begin{equation}{\label{OP3-Xn}}
x_{k,n}^\star=\left\{\begin{array}{l}{1,~\text {if}~k=k^\star=\operatorname{argmin}_{k}~\mathcal{L}_{k,n}},\\ 
{0,~\text{otherwise}.}\end{array}\right.
\end{equation}
\subsubsection{Auxiliary Variable Update}
In this part, with the newly obtained $\{\boldsymbol{f}^\star,\boldsymbol{X}^\star\}$ in the above subsections, we now try to find the optimal $\boldsymbol{\phi}^\star$, which can be solved via the following optimization problem,
\begin{subequations}\label{OP3-phi}
\begin{align}
\mathbf{P_S2}:\min_{\boldsymbol{\phi}}&\sum_{k\in\mathcal{K}}\left[\sum_{n\in\mathcal{N}_k}\frac{p_{k,n}\lambda_{k}R_{k}}{\phi_k}\!+\!\frac{\alpha_{k}\lambda_{k}R_{k}}{\phi_{k}}\!+\!\delta_{k}\phi_{k}\right]\\
\mathrm{s.t.}~&  \eqref{OP2-CT},\nonumber\\
~& 0\leq\phi_{k}\leq \tilde{r}_k, \forall k, \label{OP3-Cphi2}
\end{align}
\end{subequations}
where $\tilde{r}_k\triangleq \sum_{n\in\mathcal{N}_k} B\log_{2}(1+p_{k,n}\tilde{g}_{k,n})$. It is obvious that $\mathbf{P_S2}$ can be further decoupled into $K$ subproblems w.r.t. each user, given by
\begin{subequations}\label{OP3-phi-sub}
\begin{align}
\mathbf{P_S3}:\min_{\phi_k}&\left[\frac{\lambda_{k}R_{k}}{\phi_k}\sum_{n\in\mathcal{N}_k}p_{k,n}+\frac{\alpha_{k}\lambda_{k}R_{k}}{\phi_{k}}+\delta_{k}\phi_{k}\right]\\
\mathrm{s.t.}~& \frac{\lambda_{k}R_{k}}{\phi_{k}} + \frac{\lambda_{k}R_{k}c_{k}}{f_{k,m}}\leq T,\label{OP3-phi-sub-C1}\\
~& 0\leq\phi_{k}\leq \tilde{r}_k,\label{OP3-phi-sub-C2}
\end{align}
\end{subequations}
the optimal auxiliary variable $\boldsymbol{\phi}^\star$ can be obtained by the following theorem.
\begin{theorem}\label{theorem-phi}
Given the optimal computation capability assignment $\boldsymbol{f}^\star$, and the optimal subcarrier allocation strategy $\boldsymbol{X}^\star$, the optimal $\phi^{\star},\forall k$ is given by
\begin{equation}{\label{OP3-Xn2}}
\phi_{k}^\star\!=\!\left\{
\begin{aligned}
&{\phi_{k,1}, ~\text{if}~\phi_k^o <\phi_{k,1}},\\
&{\phi_k^o,~~~\text {if}~\phi_{k,1}\leq \phi_k^o\leq \tilde{r}_{k}}, \\ 
&{\tilde{r}_{k},
	~~~\text{otherwise}},
\end{aligned}\right.
\end{equation}
where $\phi_{k,1}$ and $\phi_k^o$ are given by
\begin{eqnarray}
&&\phi_{k,1}=\frac{\lambda_{k}R_{k}f_{k,m}}{Tf_{k,m}-\lambda_{k}R_{k}c_{k}},\\
&&\phi_k^o=\sqrt{\left(\alpha_{k}+\sum_{n\in\mathcal{N}_k}p_{k,n}\right)\tilde{\lambda}_k},
\end{eqnarray}
where $\tilde{\lambda}_k=\frac{\lambda_kR_k}{\delta_k}$. 
\end{theorem}
\begin{IEEEproof}
The proof is presented in Appendix B.
\end{IEEEproof}

%%%%%%%%%%%%%%%%%%%%%%%%%%%%%%%%%%%%%%%%%%%%%%%%%%%%%%%%%%%
\subsubsection{Lagrange Multipliers Update} In this subsection, since $\boldsymbol{f^\star}$,  $\boldsymbol{X^\star}$ and $\boldsymbol{\phi}^\star$ are obtained, we can deal with the dual problem \eqref{OP3-DP}, which is a convex function, by updating $\boldsymbol{\alpha}$, $\boldsymbol{\beta},~\boldsymbol{\delta}$ and $\gamma$ using the subgradient method.
The dual variables $(\boldsymbol\alpha,\boldsymbol\beta,\boldsymbol\delta,\gamma)$ can be updated according to the following formulations,
\begin{equation}\label{OP3-UD}
\begin{aligned}
&\alpha_{k}^{j+1}=\left[\alpha_{k}^{j}-\zeta_{k}\left
(\frac{\lambda_kR_{k}}{\phi_{k}}+\frac{\lambda_kR_kc_{k}}{f_{k,m}}-T\right)\right]^{+},
\\&\beta_{k}^{j+1}=\left[\beta_{k}^{j}-\eta_{k}\left(\sum_{n\in\mathcal{N}}x_{k,n}p_{k,n}-p_{k}^{\text{max}}\right)\right]^{+},\\
&\delta_{k}^{j+1}=\left[\delta_{k}^{j}-\xi_{k}\left(\phi_k-r_{k}\right)\right]^{+},\\
&\gamma^{j+1}=\left[\gamma^{j}-\theta\left(\sum_{k\in\mathcal{K}}f_{k,m} - F\right)\right]^{+},
\end{aligned}
\end{equation}
where $\zeta_k$, $\eta_k$, $\xi_{k}$ and $\theta$ are the stepsizes related to each dual variable during iterations, the details of which are given in TABLE \uppercase\expandafter{\romannumeral2}.

According to \ref{3A}, \ref{3B} and \ref{3C}, the details to solve $\mathbf{P}$ are summarized in Algorithm 3 as follows.
\begin{algorithm}[h]
	\begin{algorithmic}[1]
	  \caption{Proposed Algorithm}  
		\State {\bf Initialization:} given $\{\boldsymbol{p},\boldsymbol{f},\boldsymbol{X}, \boldsymbol{\phi},\boldsymbol\alpha,\boldsymbol\beta,\boldsymbol{\delta},\gamma\}$, we set $z=0$, and denote $z_{\text{max}}$ as the maximum number of iterations.
		\Repeat (from 2 to 14)
		\State Determine offloading ratio $\boldsymbol\lambda$ according to \eqref{OP-lkm}.
		\State Determine transmission power $\boldsymbol{p}$ by Algorithm 1;
			  \Repeat (from 5 to 12)
				  \Repeat  (from 6 to 10)
			  \State Allocate computing resource $\boldsymbol{f}$ by Algorithm 2; 
			  \State Determine subcarrier allocation $\boldsymbol{X}$ according to \eqref{OP3-Xn};
			  \State Update $\boldsymbol\phi^{\star}$
			  \Until Lagrangian function converges.
			\State Update $\boldsymbol\alpha$, $\boldsymbol{\beta}$, $\boldsymbol{\delta}$ and $\gamma$ resorting to \eqref{OP3-UD}; 
		%更新对偶 
			  \Until {$\boldsymbol\alpha, \boldsymbol\beta, \boldsymbol\delta, \gamma$ converge.}
			  \State $z = z + 1$
	  \Until $z>z_{\text{max}}$
	  \end{algorithmic}  
	\end{algorithm}
%%%%%%%%%%%%%%%%%%%%%%%%%第二层算法
\subsection{Complexity Analysis}
In this part, the complexity of our proposed algorithm to solve $\mathbf{P}$ is discussed. We decompose the original problem into three subproblems: $\mathbf{P_O}$, $\mathbf{P_T}$ and $\mathbf{P_S}$, then solve them iteratively, and the maximum number of iterations is $z_{\text{max}}$.

Firstly, we compute the offloading ratio $\boldsymbol\lambda$ for $\mathbf{P_O}$, and the complexity of solving $\mathbf{P_O}$ is $\mathcal{O}(K)$, a linear complexity with the number of users $K$.

Secondly, we have the three layer iterations to solve $\mathbf{P_T}$: 1) The first-layer (innermost) iteration includes $\boldsymbol{p}$ and Lagrangian variables $(\boldsymbol{\varphi},\boldsymbol{\vartheta})$, where the computing complexity to solve $\boldsymbol{p}$ according to \eqref{pkn_op} is $\mathcal{O}(KN)$ and the updating complexity of Lagrangian variables is $\mathcal{O}(K^2)$. Hence, the calculation complexity of this iterative process is $\mathcal{O}(K^2+KN)$. 2) The second-layer iteration is the updating iteration of auxiliary variables $(\boldsymbol{a},\boldsymbol{b})$, which has the complexity independent of $K$, thus the calculation complexity of this layer is $\mathcal{O}(K)$. 3) The third-layer (outermost) iteration is related to the fixed point $p_{k,n}^r$, and the computation complexity of this layer is $\mathcal{O}(KN)$. Therefore, the complexity of solving $\mathbf{P_T}$ is $\mathcal{O}(K^{3}N^2+K^{4}N)$. Since the number of subcarriers $N$ is greater than the number of users $K$, the complexity is $\mathcal{O}(K^{3}N^2)$ for $\mathbf{P_T}$.

Thirdly, we have the two layer iterations to compute $\boldsymbol{X}$ and $\boldsymbol{f}$ for $\mathbf{P_S}$. In the inner iteration, the method of solving $\boldsymbol{f}$ is dichotomy, and thus the computation complexity is $\mathcal{O}(K\log_2(F))$. Meanwhile, we get the subcarrier allocation strategy according to \eqref{OP3-Xn}, the computation complexity of which is $\mathcal{O}(KN)$, and the comuptation complexity related with $\phi$ is $\mathcal{O}(K)$. Then, we optimize $(\boldsymbol{f},\boldsymbol{X},\boldsymbol{\phi})$ iteratively until Lagrange function converges, and the computation complexity is $\mathcal{O}(K)$. In the outer iteration, the complexity of updating Lagrangian variables $(\boldsymbol\alpha,\boldsymbol\beta,\boldsymbol\delta,\gamma)$ is $\mathcal{O}(K^2)$. Therefore, the complexity of solving $\mathbf{P_S}$ is $\mathcal{O}(K^4\log_2(F)+K^4N+K^4)$, which can be simplified as $\mathcal{O}(K^{4}N)$. 

Finally, the computation complexity to solve the above three subproblems iteratively is
$\mathcal{O}(z_{\text{max}})$. 
In summary, our proposed algorithm PA has a total computation complexity $\mathcal{O}(Kz_{\text{max}}+K^{3}N^{2}z_{\text{max}}+K^{4}Nz_{\text{max}})$, which can abbreviated as $\mathcal{O}(K^{3}N^{2}z_{\text{max}})$.

\section{Equal-power-allocation based algorithm}\label{OP2-EPA}
Instead of utilizing the first-order Taylor expansion to cope with the non-convex constraint \eqref{OP-CTU} of \ref{3B}, which further makes the complexity of the algorithm in \ref{3B} unfavorable for practical application with the increase of $K$ and $N$, another algorithm is proposed to reduce complexity based on the equal power allocation (EPA) in this section.

Firstly, we make the following approximation at the high signal-to-noise ratio (SNR) regime at the right-hand side of \eqref{OP-CTU}:
\begin{equation}\label{approx}
\sum_{n\in\mathcal{N}_{k}}\log_{2}(p_{k,n}\tilde{g}_{k,n})\approx \sum_{n\in\mathcal{N}_{k}}\log_{2}(1+p_{k,n}\tilde{g}_{k,n}),
\end{equation}
where the high SNR approximation made by \eqref{approx} is commonly used to simplify the derivations and get more insight of the problem \cite{tse2005fundamentals}. The difference between the actual value and the approximate value becomes negligible as the SNR increases, and hence the constraint \eqref{OP-CTU} can be recast as
\begin{equation}\label{OP-CTU1}
\begin{aligned}
\sum_{n\in\mathcal{N}_{k}}\log_{2}(p_{k,n}\tilde{g}_{k,n})\geq \frac{\lambda_{k}R_{k}}{BT_{k,u}}, \forall k, 
\end{aligned}    
\end{equation}
where the left-hand side of \eqref{OP-CTU1} serves as a tight lower-bound for $\sum_{n\in\mathcal{N}_{k}}\log_{2}(1+p_{k,n}\tilde{g}_{k,n})$ at the high SNR regime, and the approximation is also accurate even at moderate-low SNR regime \cite{tang2007cross}. However, constraint \eqref{OP-CTU1} is still non-convex w.r.t. $p_{k,n}, \forall k,n$. In order to further deal with constraint \eqref{OP-CTU1} and make the optimization problem tractable, we assume that user $k$ will equivalently distribute its transmission power $p_k$ among its allocated subcarrier $n$, where $n\in\mathcal{N}_k$. That is to say, user $k$ will transmit the offloading data to the MEC-integrated BS using the same transmission power at each subcarrier belongs to it. With some simple transformations, \eqref{OP-CTU1} can be rewritten into a linear constraint, given by
\begin{equation}\label{OP-CTU2}
\begin{aligned}
\bar{p}_{k} \geq 2^{|\bar{N}_k|}, \forall k, 
\end{aligned}    
\end{equation}
where $\bar{p}_{k}$ is the equal power allocation for each subcarrier of user $k$, and $|\bar{N}_k|$ is given by
\begin{equation}
|\bar{N}_k|=\frac{1}{|N_k|}\left[\frac{\lambda_{k}R_{k}}{BT_{k,u}}-\sum_{n\in\mathcal{N}_{k}}\log_{2}(\tilde{g}_{k,n})\right].
\end{equation}

Then, we can reformulate $\mathbf{P_T}$ with the latest contraint \eqref{OP-CTU2} as follows,
\begin{subequations}\label{OP-PE}
	\begin{align}
	\mathbf{P_{ET}}:~\min_{\boldsymbol{\bar{p}}}~&\sum_{k\in\mathcal{K}}\bar{E}_{k,u}\\
	\mathrm{s.t.} ~& 2^{|\bar{N}_k|}\leq \bar{p}_{k}\leq\frac{p_k^\text{max}}{|N_k|},\forall k,
	\end{align}
\end{subequations}
where $\boldsymbol{\bar{p}}=\{\bar{p}_{k}\},\forall k$ and
\begin{equation}
\bar{E}_{k,u}=\frac{|N_k|\bar{p}_{k}\lambda_kR_{k}}{\sum_{n\in\mathcal{N}_k}B\log_{2}\left(1+\bar{p}_{k}\tilde{g}_{k, n}\right)}.
\end{equation}

According to the procedures described in \ref{3B}, some moderate changes should be made to adapt to the current situation before directly applied. Similar to \eqref{OP2-P2}-\eqref{OP2-b}, the Lagrangian is given as
\begin{align}
&\mathcal{L}(\boldsymbol{\bar{p}}, \boldsymbol{\varphi}, \boldsymbol{\vartheta})=\sum_{k\in\mathcal{K}}b_{k}\left[d_{k}(\bar{p}_k)-a_kh_{k}(\bar{p}_k)\right]\\
&+\sum_{k\in\mathcal{K}}\varphi_{k}\left(\bar{p}_{k}-\frac{p_k^\text{max}}{|N_k|}\right)+\sum_{k\in\mathcal{K}}\vartheta_{k}\left(2^{|\bar{N}_k|}-\bar{p}_k\right),\nonumber
\end{align}
where $d_{k}(\bar{p}_k)$ and $h_{k}(\bar{p}_k)$ are given as follows, respectively,
\begin{eqnarray}
&&d_{k}(\bar{p}_k)=|N_k|\bar{p}_{k}\lambda_kR_{k},\\
&&h_{k}(\bar{p}_k)=\sum_{n\in\mathcal{N}_k}B\log_{2}\left(1+\bar{p}_{k}\tilde{g}_{k, n}\right).
\end{eqnarray}

Referring to Theorem \ref{th1}, the optimal transmission power for those subcarriers assigned to the $k$-th user could be obtained through solving the following equation:
\begin{align}\label{pd}
\frac{\partial\mathcal{L}(\boldsymbol{\bar{p}}, \boldsymbol{\varphi}, \boldsymbol{\vartheta})}{\partial\bar{p}_k}=&b_k|N_k|\lambda_kR_k+\varphi_{k}-\vartheta_{k}\nonumber\\
&-\sum_{n\in\mathcal{N}_k}\frac{a_kb_kB\tilde{g}_{k,n}}{(1+\bar{p}_k\tilde{g}_{k,n})\ln 2}=0,
\end{align}
which can be solved by Matlab. Since the closed-form solution of $\bar{p}_k,\forall k$ is difficult to obtain, we have the following propositions to achieve the closed-form solution of $\bar{p}_k,\forall k$ based on different assumptions: 1) equivalent channel gains; 2) high SNR approximation, to get further knowledge about the relationship between the transmission power and other variables.
\begin{proposition}\label{pro2}
If the channel gains $\tilde{g}_{k,n}, n\in\mathcal{N}_k$ are equivalent for the subcarriers allocated to the $k$-th user, $\forall k$, defined by $\tilde{g}_{k}$, the optimal transmission power $\bar{p}_k^\star$ can be directly deduced from \eqref{pd}, given by
\begin{equation}\label{pro2-pk}
\bar{p}_k^\star=\left[\frac{a_kb_kB|N_k|}{\left(b_k\lambda_kR_k|N_k|+\varphi_{k}-\vartheta_{k}\right)\ln 2}-\frac{1}{\tilde{g}_k}\right]^+.
\end{equation} 
\end{proposition}
\begin{remark}
It is straightforward to know that user $k$ will assign more power $\bar{p}_k^\star$ to transmit if it has a better channel gain, and also allocate transmission power according to the assigned number of subcarriers $|N_k|$, which is different from Theorem \ref{th1}. These can be seen from the optimal formulation of $\bar{p}_k^\star$ in \eqref{pro2-pk}.
\end{remark}

\begin{proposition}\label{pro3}
In the high SNR regime, the optimal transmission power $\bar{p}_k^\star$ can be directly deduced from \eqref{pd} by assuming $(1+\bar{p}_k\tilde{g}_{k,n})\approx\bar{p}_k\tilde{g}_{k,n}$, given by
\begin{equation}
\bar{p}_k^\star=\left[\frac{a_kb_kB|N_k|}{\left(b_k\lambda_kR_k|N_k|+\varphi_{k}-\vartheta_{k}\right)\ln 2}\right]^+.
\end{equation} 
\end{proposition}
\begin{remark}
The high SNR approximation is commonly used in the literature \cite{tse2005fundamentals, tang2007cross}, as mentioned above. Different from Theorem \ref{th1} and Proposition \ref{pro2}, the transmission power $\bar{p}_k^\star$ has no relationship with the channel gain under the high SNR hypothesis similar to Corollary \ref{coro1}.
\end{remark}
% \input{Upper/up.tex}

%%%%%%%%%%%%%%%%%%%%%%%%%%%%%%%%%%%%%%%%%%%%%%%%
\section{SIMULATION AND NUMERICAL RESULTS}\label{5}
This section presents the numerical results to demonstrate the better performance of our proposed algorithm compared with the reference schemes\footnote{Since there is no similar scenario in the literature, which investgated the energy-aware offloading in time-sensitive MEC network based on OFDMA, we propose some reference schemes for comparison herein.}, named as
\begin{itemize}
    \item \textbf {Local computation algorithm (LC)}, where all user devices process their own tasks by local CPU.
    \item \textbf {Fixed offloading ratio algorithm (FR)}, where we assign fixed offloading ratios for each user satisfying the time constraints \eqref{OP1-C2}.
\end{itemize}

In the simulations, the path loss model is Rayleigh distributed and denoted by $|\beta|d^{-2}$, where $|\beta|$ and $d$ represent the short-term channel fading and the distance between two nodes, respectively. User devices have the same maximum transmit power, which are evenly and independently distributed in a circular area around the MEC server with a radius of 50 meters. Moreover, we set $(\kappa_k,\kappa_m)$ as $(10^{-24},10^{-26})$, and other parameters employed in the simulations are summarized in TABLE \uppercase\expandafter{\romannumeral2}, unless otherwise mentioned. 
\begin{table}  
\caption{SIMULATION PARAMETERS}  
\begin{tabularx}{9cm}{lXl}  
\hline  
\bf MEC System Parameters   &   \bf Values \\  
\hline  
The CPU frequency of MEC sever  &  10GHz \\
The CPU frequency of mobile users & 0.6-0.7GHz\\
The transmit power of users        & 30dBm\\
Input data size of users          & $10^3-1.5\times 10^3$ bits\\
Maximum accomplished deadline $T$   &45ms \\ 
The computation workload/intensity $c_{k}$ & $1000 -1100$ cycles/bit\\
Background noise $\sigma^{2}$     &$10^{-13}$W\\
Subcarrier bandwidth $B$            &12.5KHz\\
\hline  
\bf Lagrange Iteration Parameters   &   \bf Values \\  
\hline  
Maximum number of iterations $z_\text{max}$ &600\\
Iterations precision & $10^{-5}$\\
Stepsizes $\zeta_k$           & $10^{-6}$\\
Stepsizes $\eta_{k}$           & $10^{-8}$\\
Stepsizes $\xi_{k}$				&$10^{-8}$\\
Stepsize $\theta$           & $10^{-15}$\\
Stepsize $\iota_{k}$			&$10^{-4}$\\
Stepsize $\nu_{k}$			&$10^{-8}$\\
\hline  
\end{tabularx}\vspace{-0.85em}
\end{table}

\begin{figure}  
	\centering  
	\includegraphics[width=8.5cm]{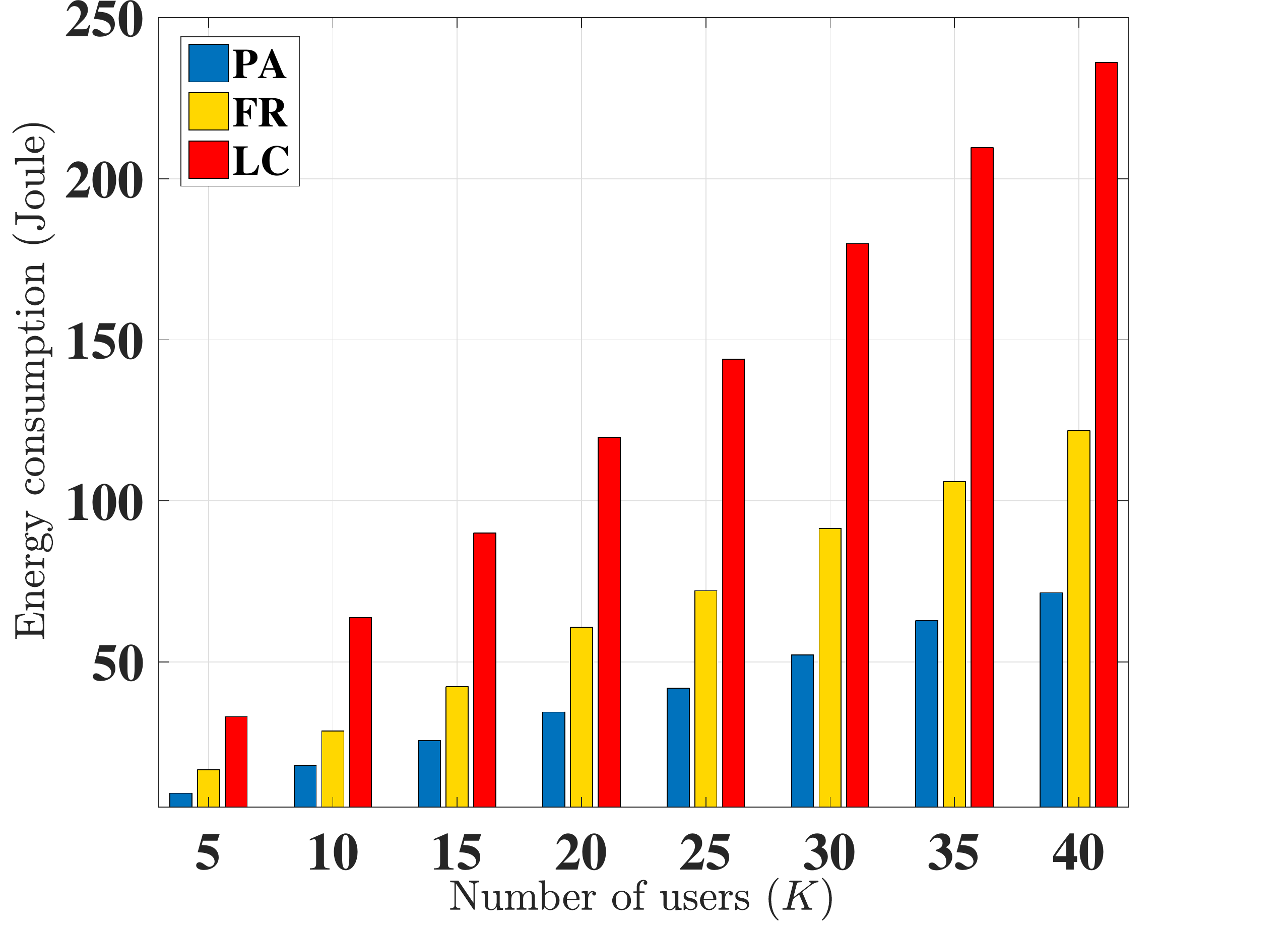}\\
	\caption{ The comparison of the total energy consumption versus different numbers of users where $N= 512$ for different algorithms.}  
	\label{fig_ene} 
\end{figure}

In Fig.\ref{fig_ene}, for the proposed and reference schemes, we plot the total energy consumption according to different numbers of users where $N=512$. With the growth of the number of users, the total energy consumption of different algorithms are increasing, and our proposed algorithm (PA) can help save 20\% -- 50\% and 40\% -- 70\% energy consumption compared with FR and LC.

\begin{figure}  
	\centering  
	\includegraphics[width=8.5cm]{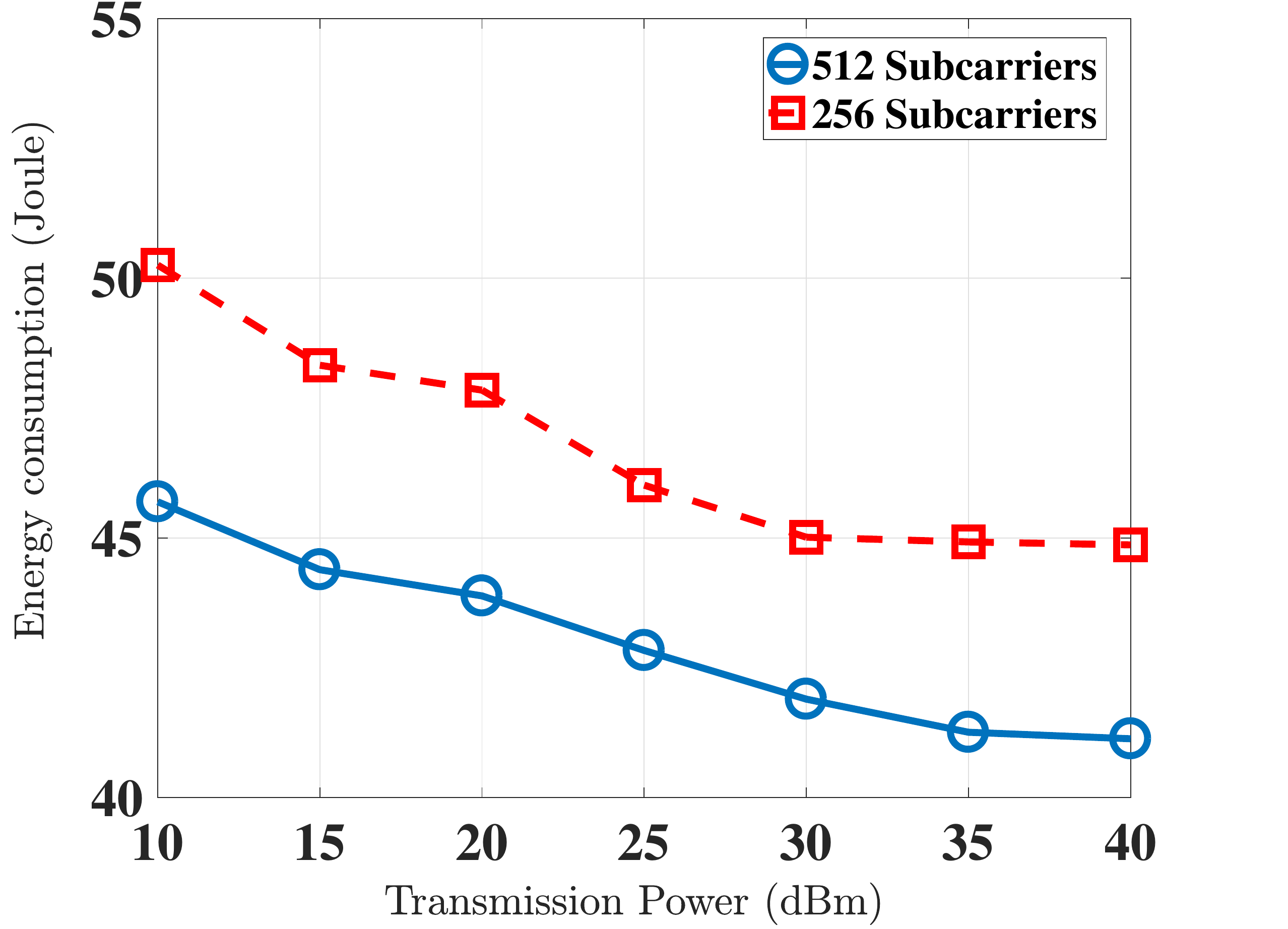}\\
	\caption{The total energy consumption of versus different transmission power of users, where $K=25$.}  
	\label{fig_power}
\end{figure}

In Fig.\ref{fig_power}, we investigate how the transmission power of users affects the energy consumption of MEC network in our considered scnerio, where $K=25$. It is observed that the energy consumption of network is decreasing w.r.t. the increasing transmission power of users. The reason is that users' transmission rate will grow with the increment of the corresponding transmission power, and further reduce the uplink transmission time $t_{k,u}$. Therefore, users can offload more data to the MEC-integrated BS within the same period of time, which helps reduce the energy consumption with more efficient computation capability at MEC. However, since the impact of uplink transmission time on the energy consumption is relatively small compared to the execution time, the reducing extent of energy consumption will gradually decrease when the transmission power increases to a certain range.
\begin{figure}  
\centering  
\includegraphics[width=8.5cm]{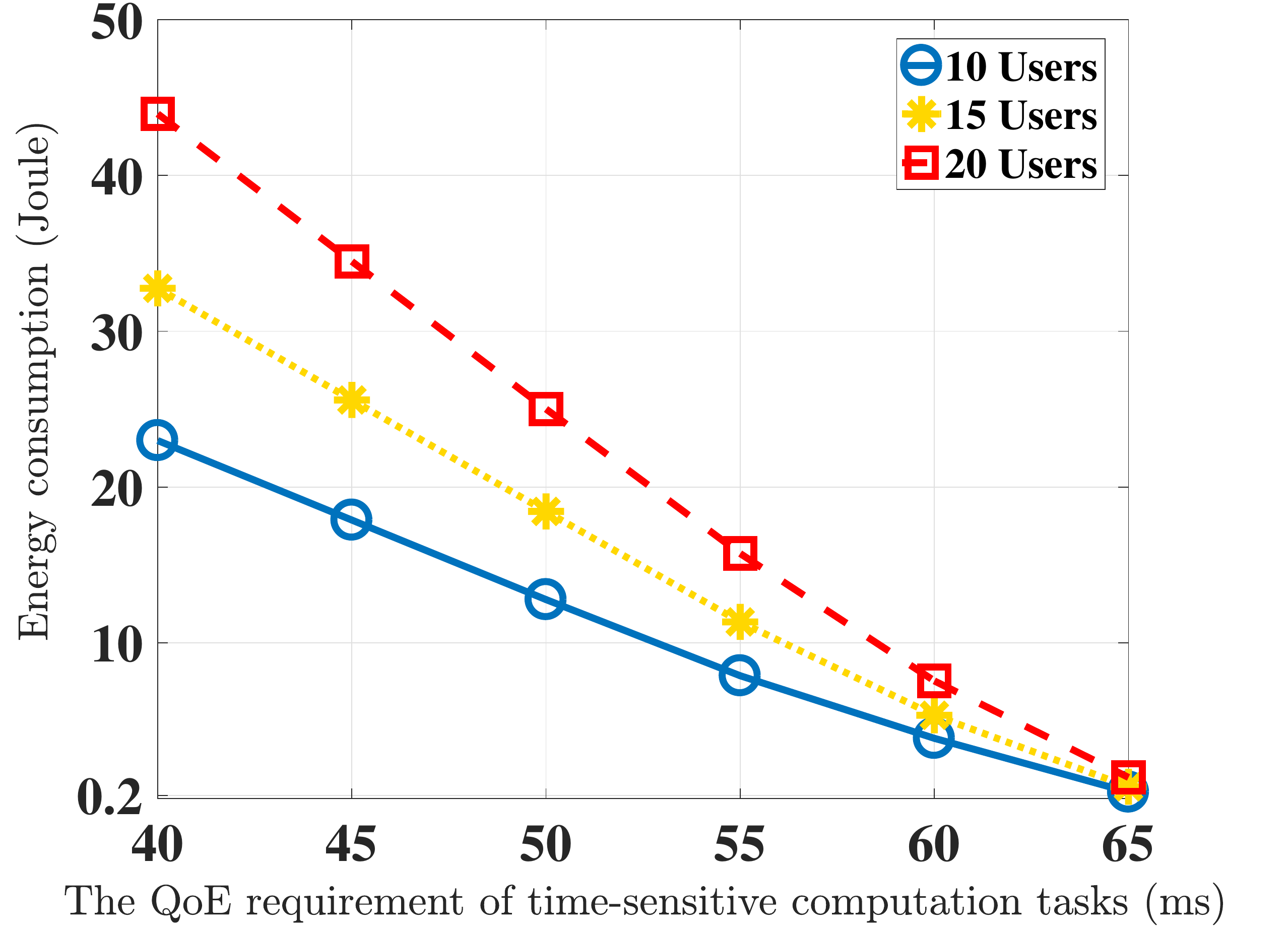}\\
\caption{The total energy consumption of versus different the QoE requirements of time-sensitive computation tasks of users where $N=512$.}  
\label{fig_time}
\end{figure}

In Fig.\ref{fig_time}, we study the influence of the QoE requirements of time-sensitive computation tasks of users on the total energy consumption for different numbers of users.
On the observation of Fig.\ref{fig_time}, the total energy consumption is reducing along with the gradually decreasing QoE requirement of time-sensitive computation tasks. This is because users can ask for more help from MEC server by uploading more data since the requirement of latency is not so strict, and thus users can reduce more energy dissipation. Moreover, the reducing extent of energy consumption w.r.t. the latency will gradually decrease when the requirement of latency approaches to a certain threshold, that is to say, the growing latency has a diminishing influnce on the total energy consumption.
\begin{figure}  
	\centering  
	\includegraphics[width=8.5cm]{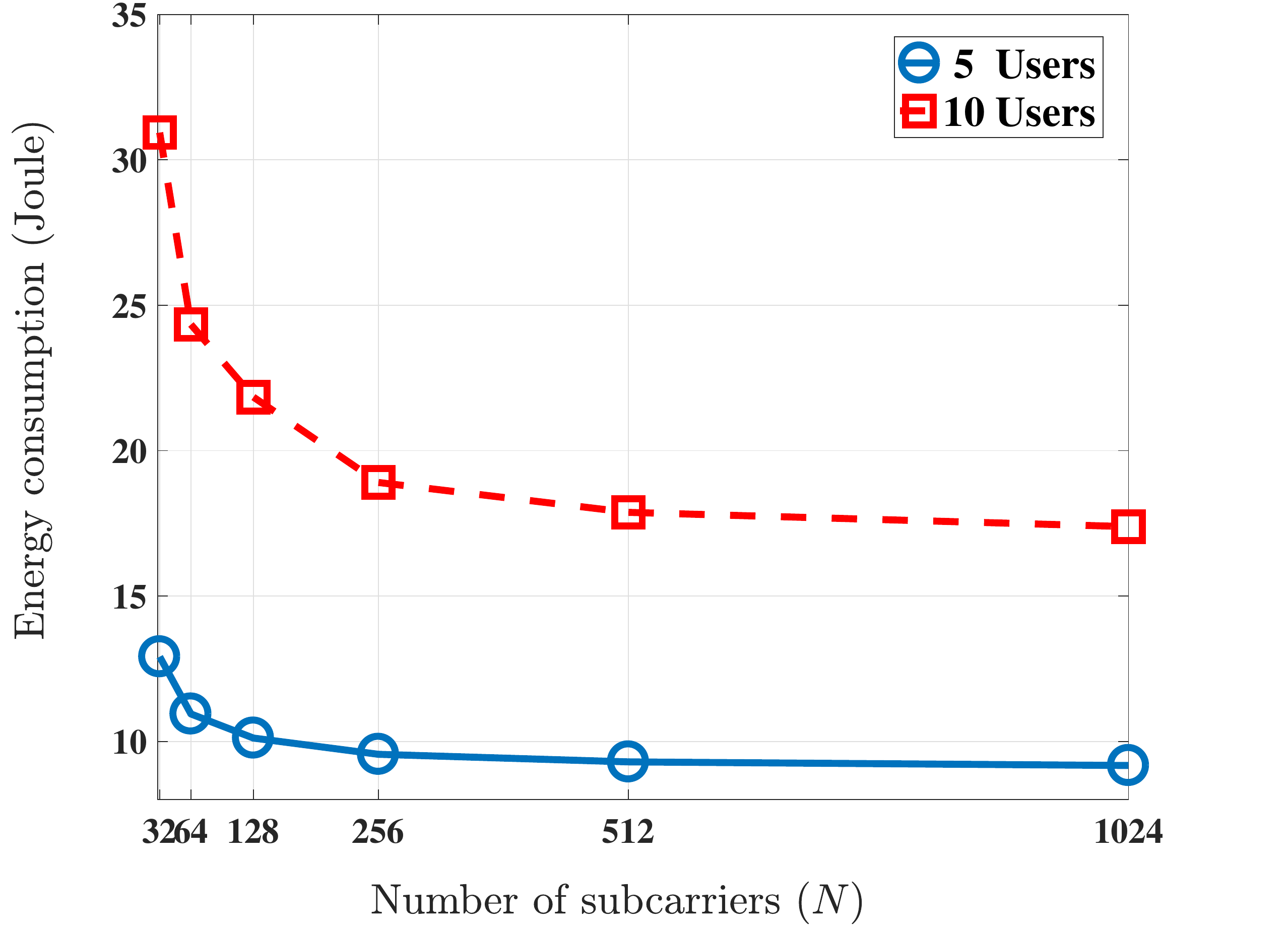}\\
	\caption{The comparison of the total energy consumption versus different numbers of subcarriers.}  
	\label{fig_sub}
\end{figure}

In Fig.\ref{fig_sub}, we evaluate the effect of the number of subcarriers on the total energy consumption. Moreover, it can be seen that the total energy consumption is reducing along with the increasing number of subcarriers. The reason is that the user will have a better chance to select the subcarrier with preferable channel gain to get the reduction of transmission power in turn, meanwhile, the user can offload more data with the same amount of transmission power and thus reduce the energy consumption with the aid of MEC server. 
\begin{figure}  
	\centering  
	\includegraphics[width=8.5cm]{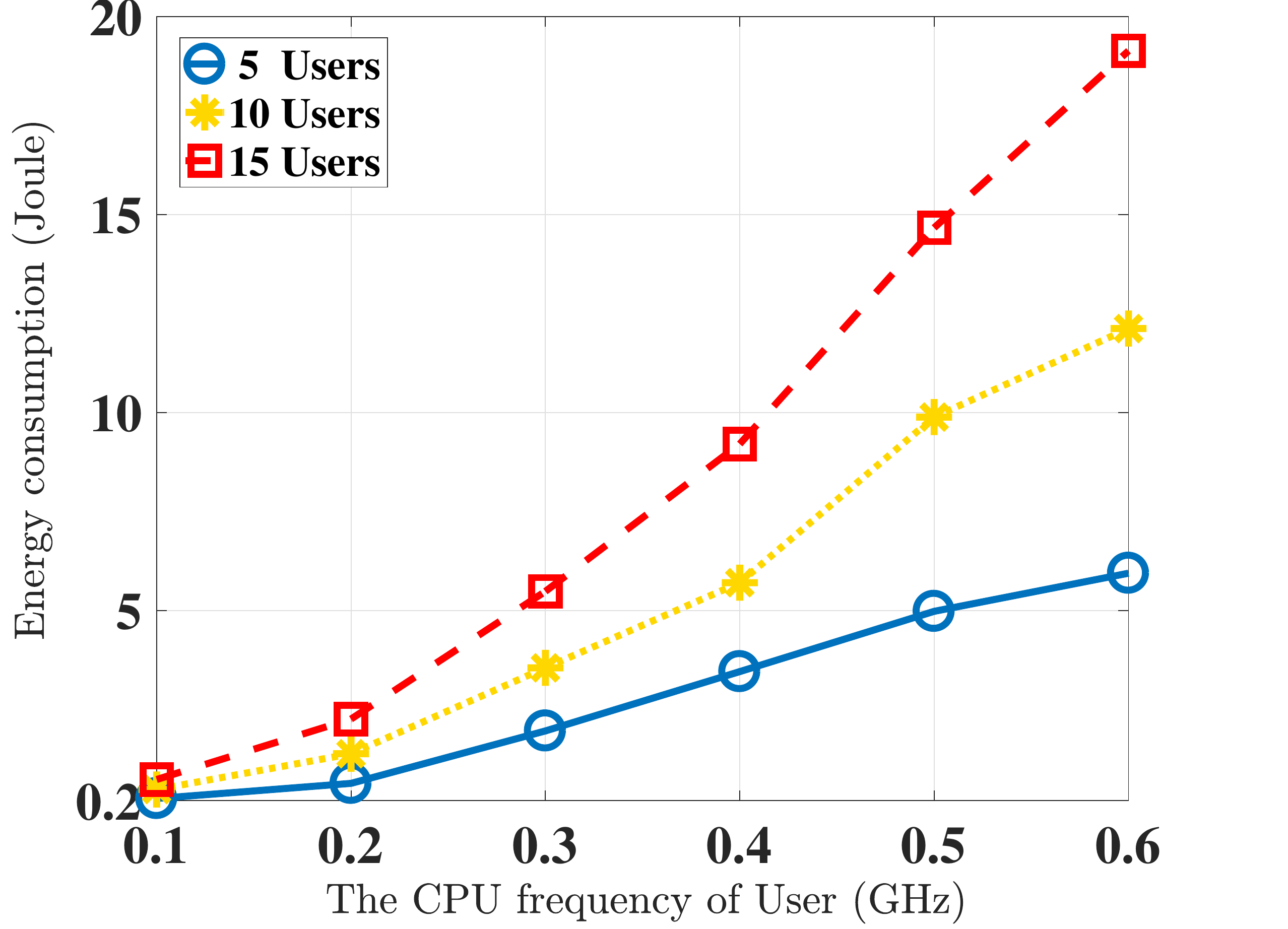}\\
	\caption{The comparison of the total energy consumption versus different CPU frequencies of users where $N=512$.}  
	\label{fig_fk} 
\end{figure}

In Fig.\ref{fig_fk}, we present the performances of the total energy consumption versus different CPU frequencies of users for $K = 5, 10, 15$. It is noted that the total energy consumption is growing in the wake of the increment of CPU frequencies of users. The reason is that users prefer to compute data locally along with their enhanced computation capabilities according to Remark \ref{remark1}, and thus the proportion of data offloading to MEC server is decreasing, which is more energy-consuming. What is noteworthy is that the increase of the total energy consumption has been slowing or saturated when the CPU frequencies of users grow in the regime of $f_k\geq0.5$GHz,  $\forall k$. This is because users are energy hogs compared with MEC server as $\kappa_k>\kappa_m$, and users are willing to offload more data to reduce their energy consumption even if they have growing computation capabilites, which has been explained in details in Remark \ref{remark2}.
\begin{figure}  
	\centering  
	\includegraphics[width=8.5cm]{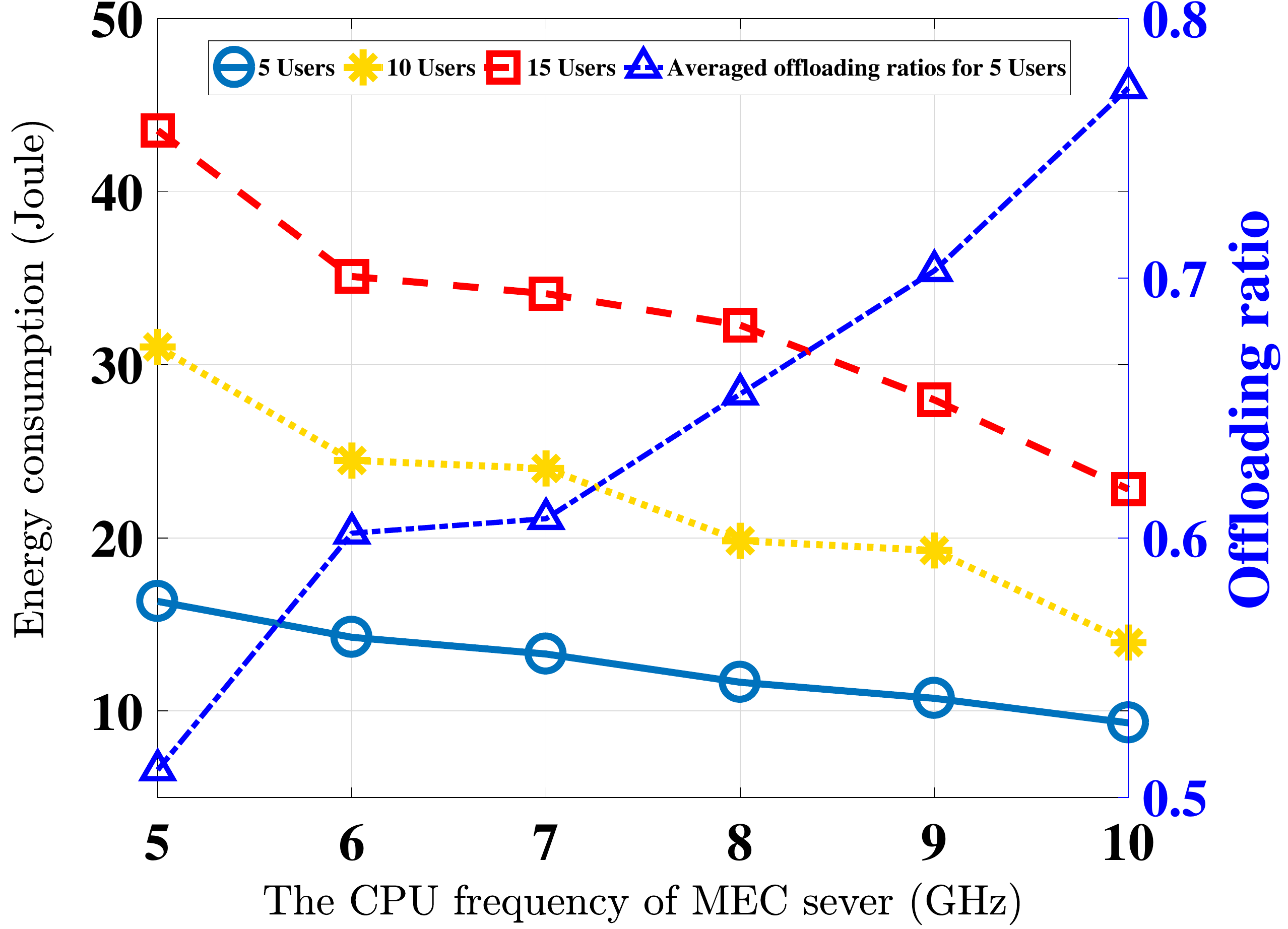}\\
	\caption{The comparisons of the total energy consumption (and the averaged offloading ratios where $K=5$) versus different CPU frequencies of MEC server where $N=512$.}  
	\label{fig_fkm} 
\end{figure}

In Fig.\ref{fig_fkm}, we show the performances of the total energy consumption versus different CPU frequencies of MEC for different numbers of users. Along with the increasing CPU frequency of MEC server, the total energy consumption of network is decreasing. The reason is that when the computation capability of MEC is strengthened, users are more inclined to offload data to MEC instead of computing locally according to Theorem \ref{theorem1}. Thus, the energy consumption will be reduced compared with local computing due to the highly efficient computation at MEC. Additionally, observed from the curve of the averaged offloading ratio for $K = 5$ in Fig.\ref{fig_fkm}, it can be seen that the uploading proportion increases with the growth of CPU frequency at MEC.
\section{CONCLUSIONS}\label{6}
In the paper, we investigate the OFDMA-based MEC network with computation capability, and design PA method to minimize the total energy consumpton of the network with the QoE requirement of time-sensitive computation tasks for users. Due to the coupled optimization variables, we decompose the formulated minimization problem into three subproblems named as $\mathbf{P_O}$, $\mathbf{P_T}$ and $\mathbf{P_S}$, and optimize them sequentially instead of solving the original problem directly. For $\mathbf{P_O}$, we can derive the closed-form solution of offloading ratios. Since the objective function of $\mathbf{P_T}$ is in the form of sum of ratios and thus difficult to cope with, we apply the equivalent parametric convex programming to optimize the transmission power of user on its assigned subcarriers with the help of auxiliary variables. As $\mathbf{P_S}$ is a challenging non-convex MINLP problem with coupled parameters, we distribute computation capability of MEC server, and allocate subcarriers to each user in an alternating manner in the dual domain. Simulation results demonstrate that the proposed algorithm can effectively reduce the total energy consumption of the network, and outperform the reference schemes with a better performance.
\appendices
\def\thesection{\Alph{section}}%
\def\thesectiondis{\Alph{section}}%
\section{Proof of Theorem 2}
\setcounter{equation}{0}
\renewcommand{\theequation}{A.\arabic{equation}}
Taking the derivative of the Lagrangian $\mathcal{L}(p_{k,n},\varphi_{k},\vartheta_{k})$ w.r.t. $p_{k,n}$ yields\\
\begin{align}\label{OP-de}
&\frac{\partial\mathcal{L}(p_{k,n},\varphi_{k},\vartheta_{k})}{\partial p_{k,n}}\\
&=b_{k}\lambda_{k}R_{k}\!-\!\frac{a_{k}b_{k}B\tilde{g}_{k,n}}{(1+p_{k,n}\tilde{g}_{k,n})\ln2}\!+\!\varphi_{k}\!-\!\frac{\vartheta_k\tilde{g}_{k,n}}{\left(1+p_{k,n}^r\tilde{g}_{k,n}\right)\ln 2}.\nonumber
\end{align}
Let $\frac{\partial\mathcal{L}(p_{k,n},\varphi_{k},\vartheta_{k})}{\partial p_{k,n}}=0$, the optimal $p_{k,n}^\star$ can be obtained, and thus we have Theorem 1.

\def\thesection{\Alph{section}}%
\def\thesectiondis{\Alph{section}}%
\section{Proof of Theorem 3}
\setcounter{equation}{0}
\renewcommand{\theequation}{B.\arabic{equation}}
We can get the following inequality about $\phi_k$ from \eqref{OP3-phi-sub-C1} and \eqref{OP3-phi-sub-C2}
\begin{equation}\label{OP3-phic3}
\begin{aligned}
\frac{\lambda_{k}R_{k}f_{k,m}}{Tf_{k,m}-\lambda_{k}R_{k}c_{k}}\leq \phi \leq \tilde{r}_k,
\end{aligned}    
\end{equation}
where $\tilde{r}_k\triangleq \sum_{n\in\mathcal{N}_k} B\log_{2}(1+p_{k,n}\tilde{g}_{k,n})$. Thus, we can rewrite $\mathbf{P_{R}3}$ as 
\begin{equation}\label{OP3-PR3A}
\begin{aligned}
\mathbf{P_R3A}:~\min_{\phi_k}~&\mathcal{T}(\phi_k)\\
\mathrm{s.t.}~& \eqref{OP3-phic3},
\end{aligned}    
\end{equation}
where $\mathcal{T}(\phi_k)=\frac{\lambda_{k}R_{k}}{\phi_k}\sum_{n\in\mathcal{N}_k}p_{k,n}+\frac{\alpha_{k}\lambda_{k}R_{k}}{\phi_{k}}+\delta_{k}\phi_{k}$, and $\mathbf{P_R3A}$ is convex since the second derivative of $\mathcal{T}(\phi_k)$ with respect to $\phi_k$ is positive.
Therefore, resorting to the first-order condition
\begin{equation}\label{OP3-dphi}
\begin{aligned}
\frac{\partial{\mathcal{T}(\phi_k)}}{\partial{\phi_k}}\!=\!\delta_k-\frac{\lambda_kR_{k}(\sum_{n\in\mathcal{N}_{k}}p_{k,n}+\alpha_k)}{\phi_k^{2}},
\end{aligned}    
\end{equation}
and letting $\frac{\partial{\mathcal{T}(\phi_k)}}{\partial{\phi_k}}=0$, the optimal $\phi_k^{o}$ is given as
\begin{equation}\label{OP3-phistar}
\begin{aligned}
\phi_k^{o}=\sqrt{\left(\alpha_{k}+\sum_{n\in\mathcal{N}_k}p_{k,n}\right)\frac{\lambda_{k} R_{k}}{\delta_k}}. 
\end{aligned} 
\end{equation}

Therefore, we can discuss the relationship between the boundaries of $\eqref{OP3-phic3}$ and $\phi_k^{o}$ to get the optimal $\phi_k^{\star}$, and obtain Theorem 2.

\def\thesection{\Alph{section}}%
\def\thesectiondis{\Alph{section}}%
\section{Proof of Proposition 1}
\setcounter{equation}{0}
\renewcommand{\theequation}{C.\arabic{equation}}
The \emph{Hessian} of $\mathcal{L}$ in \eqref{OP3-L} w.r.t. $f_{k,m}$ is given by
\begin{equation}
\frac{\partial^2\mathcal{L}(\boldsymbol{f},\boldsymbol{X},\boldsymbol\phi,\boldsymbol\alpha,\boldsymbol\beta,\boldsymbol\delta,\gamma)}{\partial{f^2_{k,m}}}=2\kappa_m\lambda_kR_kc_k+\frac{\alpha_k\lambda_kR_kc_k}{f_{k,m}^3},
\end{equation}
which is non-negative w.r.t. $f_{k,m}$, thus $\mathcal{L}$ is a convex function of $f_{k,m}$ \cite{boyd2004convex}.
%\printbibliography
\bibliographystyle{IEEEtran}
\bibliography{main.bib}
\end{document}